\RequirePackage{fix-cm}
\documentclass{svjour3}                     % onecolumn (standard format)
\usepackage{etoolbox}
\makeatletter
\def\makeheadbox{}   % remove "Noname manuscript No. (will be inserted by the editor)"
\patchcmd{\@maketitle}{\hrule\@height0.35mm}{}{}{}
\makeatother

\smartqed  % flush right qed marks, e.g. at end of proof
\usepackage{graphicx}
\usepackage{orcidlink}
\usepackage{xcolor}
\usepackage{lmodern} 
\usepackage{amsmath}
\usepackage{amssymb}
\usepackage{array}
\usepackage[most]{tcolorbox}
\usepackage{multirow}
\usepackage[normalem]{ulem}
\usepackage{subcaption}
\usepackage{booktabs}
\usepackage{fontawesome}
\usepackage{listings}
\usepackage[T1]{fontenc}
\usepackage{inconsolata} % better monospace font
\usepackage{caption}
\usepackage{upquote}
\usepackage{textcomp}

\newcommand{\ibq}[1]{\textsuperscript{\scriptsize(#1)}}

\definecolor{codebg}{RGB}{248,248,248}
\definecolor{keywordcolor}{RGB}{33,74,135}
\definecolor{commentcolor}{RGB}{0,120,0}
\definecolor{stringcolor}{RGB}{163,21,21}
\definecolor{framecolor}{RGB}{220,220,220}

\lstdefinestyle{pythonstyle}{
    language=Python,
    backgroundcolor=\color{codebg},
    basicstyle=\footnotesize\ttfamily,
    keywordstyle=\color{keywordcolor}\bfseries,
    commentstyle=\color{commentcolor}\itshape,
    stringstyle=\color{stringcolor},
    showstringspaces=false,
    breaklines=true,
    breakatwhitespace=true,
    frame=single,
    rulecolor=\color{framecolor},
    framesep=6pt,
    frameround=tttt,
    tabsize=4,
    columns=fullflexible,
    keepspaces=true,
    xleftmargin=0.5em,
    framexleftmargin=0.5em,
    aboveskip=0.8em,
    belowskip=0.8em,
    upquote=true
}

\begin{document}

\title{Do Influence Tactics Matter? Investigating Prompt Framing Effects in LLM Code Generation}

\titlerunning{Do Influence Tactics Matter?}        % if too long for running head

\author{Alex Deaconu\textsuperscript{*} \orcidlink{0009-0003-6810-9619} \thanks{\textsuperscript{*} Alex Deaconu and Anubhav Gupta shared first authorship.}\and
        Anubhav Gupta\textsuperscript{*} \orcidlink{0000-0003-0187-8192} \and
        Manaal Basha \and
        Nicholas Haydu \and
        Gema Rodr\'iguez-P\'erez\orcidlink{0000-0002-0062-8418} }

\institute{ Extended author information available on the last page of the article
}

\date{}

\maketitle
\noindent\footnotesize
This version of the article has been accepted for
publication, after peer review, but is not the Version of Record
and does not reflect post-acceptance improvements or any
corrections. The Version of Record is available online at:
\url{http://dx.doi.org/10.1007/s10664-026-10934-z}. Use of this
Accepted Version is subject to the publisher's Accepted Manuscript
terms of use.
\\
\normalsize

\begin{abstract}
Large Language Models (LLMs) are increasingly integrated into software engineering workflows, helping developers write, debug, test, and maintain code. While prompt wording and structure are known to influence model performance, the impact of psychologically inspired prompt framings remains unexplored. This study investigates whether different psychology-based communication strategies that humans use to persuade or motivate others can lead to more effective prompt framing, which may, in turn, affect LLM behaviour in coding tasks. Drawing on Yukl \& Falbe's well-known taxonomy, we operationalized eight influence tactics (like rational persuasion, ingratiation, and exchange) into reproducible prompt templates. These prompt templates were evaluated across five leading open-weight LLMs using two widely adopted benchmarks: LiveCodeBench and SWE-bench Verified. We assessed the resulting code output on four key software quality dimensions: functional correctness, quality, maintainability, and security. Our results show that certain influence-induced prompt framings, particularly those emphasizing urgency, were associated with reduced correctness and security. This work presents the first large-scale empirical study of influence-induced prompt framing in software engineering tasks, offering insights into how linguistic cues may shape LLM outputs. We conclude with practical insights for designing transparent and interpretable human-AI interactions in code generation.

\end{abstract}

\section{Introduction}
\label{intro}
Large Language Models (LLMs) have rapidly emerged as influential tools in Software Engineering (SE), supporting a variety of tasks such as code synthesis~\cite{rasheed2025large}, API documentation~\cite{deng2025lrasgen}, automated bug repair~\cite{zhang2025can}, and supporting testing workflows~\cite{zhang2025large}. As LLMs become more integrated into professional development workflows, recent studies suggests that the way \textit{developers phrase and frame their prompts} can influence model behaviour and generated code quality~\cite{fagadau2024analyzing,white2023prompt}. For instance, Tony et al.~\cite{tony2024prompting} found that the \textit{Recursive Criticism} and \textit{Improvement} technique reduced security weaknesses across some LLMs. Similarly, Liu et al.~\cite{liu2023improving} showed that chain-of-thought prompt designs notably improve ChatGPT's coding performance.

Prior work has explored syntactic prompt engineering~\cite{chen2021evaluating}, reasoning-based strategies like chain-of-thought~\cite{wei2022chain,liu2023improving}, and the role of explicit instructions in improving code generation~\cite{bsharat2023principled}. In contrast, research in organizational psychology has established that variations in request framing can substantially impact human compliance and performance~\cite{flusberg2024psychology}. For example, Yukl \& Falbe's influential taxonomy of influence tactics~\cite{yukl1990influence} shows how subtle shifts in approach, such as appeals to rational arguments, expressions of urgency, or offers of reciprocity, can change outcomes in collaborative settings. Translating this idea to software engineering, this paper investigates: \textit{To what extent do prompt framings inspired by these influence tactics lead to measurable differences in LLM-generated code?} 

This question is particularly relevant as SE increasingly relies on interactive human-AI workflows where developers iteratively refine prompts to achieve reliable, maintainable, and secure code. Human developers naturally tend to use rational persuasion to justify design decisions, invoke pressure to communicate urgency under tight deadlines, or suggest exchanges to motivate cooperation~\cite{jordan2014designing}. Understanding how \textit{psychologically inspired influence tactics} in prompt design, particularly those shaping interpersonal communication, influence model outputs is important as it can improve prompt design and reveal potential risks when AI-assisted development tools are used in real-world SE workflows.

Recent research has shown that subtle differences in tone and wording, such as politeness~\cite{quan2025human}, can alter the length and affective tone of responses from GPT-4, with more polite prompts leading to longer and more positive outputs. Yin et al.~\cite{yin2024should} observed that medium to high politeness levels correspond to fewer comprehension errors. In adversarial contexts, socially engineered prompts that mimic persuasive or deceptive framing have been found to increase the success rate of prompt injection attacks~\cite{eccws2025psychologicalPromptInjection}. These findings suggest that linguistic cues associated with human communication patterns may shape how LLMs retrieve and compose information. Because LLMs are trained on large corpora of human-authored communication, they may learn statistical associations between particular pragmatic framings and particular response styles. Under this interpretation, psychologically inspired prompt framings act as distributional cues that may steer generation behaviour without requiring the model to understand persuasion or social intent in a human sense. This motivates our investigation into whether influence-tactic framings produce measurable differences in code generation outcomes.

In this paper, we present the first empirical study examining how prompt framings inspired by psychological influence tactics affect LLM outputs in code generation tasks. Drawing on Yukl \& Falbe's~\cite{yukl1990influence} taxonomy of psychological influence strategies, we operationalized eight tactics into reproducible prompt templates. For each tactic, we used items from the Influence Behavior Questionnaire-General (IBQ-G), a validated instrument for identifying the usage of influence tactics, as requirements to build prompt templates. We evaluated these tactics across five open-weight LLMs using two benchmarks: LiveCodeBench and SWE-bench Verified. We assessed each model's outputs on four key dimensions central to software code quality: (1) functional correctness; (2) code quality; (3) maintainability; and (4) security attributes. To complement our quantitative analysis, we conducted a triangulation study to explore the qualitative differences underlying these effects and gain deeper insights into the reasons behind them.

Our results indicate that certain tactic-influenced framings, such as Pressure, were associated with reduced correctness and security. Beyond quantitative metrics, these tactic-influenced prompt framings lead to qualitative differences in the tone, structure, and reliability of model responses, with some prompting more explanation or technical detail, while others increased hallucination rates.

This paper contributes to empirical software engineering in three ways:
\begin{itemize}
    \item \textbf{Empirical Analysis:} The first large-scale study quantifying the impact of psychologically inspired prompt framings on LLM-generated code, covering over 123,000 generations for LiveCodeBench and nearly 57,000 generations for SWE-bench Verified across multiple models, runs, and influence tactics.
    %\textcolor{red}{A framework of prompting requirements, including a taxonomy and reusable templates, enabling systematic application of psychological influence tactics in LLM-driven code generation.}
    \item \textbf{Prompt design resource:} A taxonomy-aligned set of reproducible prompt templates operationalizing eight distinct influence tactics for programming and software engineering tasks. %\textcolor{red}{First empirical study examining how psychological influence tactics embedded in prompts affect the outputs of code generation tools.}
    \item \textbf{Practical insights:} Evidence-based guidance on how linguistic framing choices affect functional, maintainability, and security dimensions of AI-generated code, informing more transparent and interpretable human-AI interactions in SE. %\textcolor{red}{Insights into how influence tactics affect software quality, helping developers craft prompts more effectively.}
\end{itemize}

\section{Background and Related Work}
\label{Background}

\subsection{Influence Tactics and Psychological Framing}
In human organizational settings, an influence tactic is an approach by an agent to achieve some goal from another individual, called the target, where success depends on the tactic applied~\cite{yukl2008validation}. The original taxonomy of influence tactics was developed by Kipnis et al.~\cite{kipnis_intraorganizational_1980} across two studies. The first study asked 165 lower-level managers to write descriptions of the successful influence attempts they made towards their superiors, co-workers, or subordinates to achieve a goal. From the descriptions, 370 tactics were extracted and sorted into one of 14 greater categories. The initial categories exhibited substantial overlap, which prompted a subsequent study to determine the underlying structure of the previous tactics. The follow-up study used factor analysis on a 58-item questionnaire to refine these into eight core dimensions of influence tactics.

Yukl \& Falbe~\cite{yukl1990influence} refined the category definitions from Kipnis et al.~\cite{kipnis_intraorganizational_1980} by developing the Influence Behaviour Questionnaire (IBQ), which incorporated data from both agents and targets of influence tactics, leading to an updated nine tactic taxonomy. Yukl \& Falbe note that pressure tactics, characterized by demands or threats, are frequently employed in downward influence attempts, though organizational literature suggests such coercive approaches may provoke resistance or reduce effectiveness~\cite{yukl1990influence}. Yukl \& Seifert~\cite{yukl2002preliminary} introduced two more influence tactics, leading to a final taxonomy of 11 influence tactics. The addition of the new tactics was formally validated by Yukl, Seifert \& Chavez~\cite{yukl2005assessing} through multiple methods, including confirmatory factor analysis, which established the distinctiveness and utility of the expanded autonomy. 

The 11-tactic framework and the associated Influence Behaviour Questionnaire-General (IBQ-G) survey were analyzed in~\cite{yukl2008validation} against competing taxonomies. Confirmatory factor analysis again supported the 11-factor structure, with convergent validity tested against competing taxonomies and discriminant validity via low inter-tactic correlations. Their paper also provided the refined IBQ-G, which we utilize in the design of our influence tactic prompt template. See Table~\ref{tab:influence-tactics} for a summary of how the set of influence tactics has evolved across key studies.

\begin{table}[ht]
\small
\centering
\caption{Evolution of influence tactics across key studies.}
\begin{tabular}{l|l|l}
%\hline
\textbf{Kipnis et al.~\cite{kipnis_intraorganizational_1980}} & \textbf{Yukl \& Falbe~\cite{yukl1990influence}} & \textbf{Yukl et al.~\cite{yukl2005assessing}} \\
\hline
\hline

1. Assertiveness           & 1. Pressure                & 1. Pressure                \\
2. Ingratiation            & 2. Ingratiation            & 2. Ingratiation            \\
3. Rationality             & 3. Rational persuasion     & 3. Rational persuasion     \\
4. Sanctions               & 4. Legitimating tactics    & 4. Legitimating tactics    \\
5. Exchange                & 5. Exchange                & 5. Exchange                \\
6. Upward appeals          & 6. Personal appeals        & 6. Personal appeals        \\
7. Blocking                & 7. Coalition tactics       & 7. Coalition tactics       \\
8. Coalitions              & 8. Inspirational appeals   & 8. Inspirational appeals   \\
                        & 9. Consultation            & 9. Consultation            \\
                        &                         & 10. Apprising               \\
                        &                         & 11. Collaboration           \\

\hline
\hline

\end{tabular}
\label{tab:influence-tactics}
\end{table}

In a comprehensive meta-analysis of 49 independent samples, Lee et al.~\cite{lee2017get} found that the categories of the final influence tactic framework~\cite{yukl2002preliminary} differ in both their likelihood of success and the quality of work produced. Rational persuasion, inspirational appeal, apprising, collaboration, ingratiation and consultation all showed significant positive relationships with task-oriented outcomes. Meanwhile, the effects from coalition and exchange tactics were insignificant, and pressure tactics were reliably counter-productive to task outcomes. Besides influence tactics, other studies corroborate that work performance depends on interpersonal communication and the work environment. One study finds that toxic workplace environments (including bullying, harassment and ostracism) can adversely affect employees' work performance ~\cite{rasool2021toxic}. Additionally, evidence suggests that exposure to rudeness in email communication can lead to a decline in subsequent task performance, as demonstrated by McCarthy et al.~\cite{mccarthy2020you}. Given that LLMs are trained on a large corpora of human communication, it is reasonable to ask whether these psychologically derived tactics might also shape LLM outputs.

\subsection{Large Language Models for Code Generation}
Recent research has critically examined the reliability, accuracy, and security of code generated by LLMs. While LLMs show promise in automating coding tasks and improving developer productivity, studies have revealed persistent issues in code quality. For instance, Mohammadi Esfahani et al.~\cite{esfahani2024understanding} identified 367 defects in LLM-generated code, primarily related to functionality and algorithmic logic. However, structured prompting was found to reduce some errors. Li and Wang~\cite{zhong2024can} reported that 62\% of GPT-4-generated code contained API misuses, raising concerns about its deployment in real-world applications. For instance, a common failure pattern involved file-handling code that omitted required exception handling, such as calling \texttt{File.createNewFile()} without a 
\texttt{try-catch} block or checking \texttt{File.exists()} beforehand, producing code that is syntactically correct and functionally aligned 
with the user's intent, yet prone to crashes in production when the file 
already exists, or the parent directory is missing. Beer et al.~\cite{beer2024examination} observed significant variability in code correctness and quality between models like ChatGPT and Copilot across different programming languages and time periods. Security remains a major concern, particularly due to LLMs being trained on potentially vulnerable public code repositories~\cite{mohsin2024can}. In response, researchers have proposed secure behavioural learning frameworks and rigorous security evaluations. Complementing these technical evaluations, Rasheed et al.~\cite{rasheed2025large} surveyed 60 software practitioners to assess LLMs' usability and performance in practical settings. Together, these findings underscore the importance of continued research to enhance the robustness, security, and real-world applicability of LLM-generated code.

\subsection{Prompt Engineering}
Earlier LLMs have demonstrated that carefully crafted prompts could induce strong performance on new tasks without fine-tuning. For example, in-context learning, which provides a few input-output examples in the prompt, was introduced with GPT-3~\cite{brown2020language}. In-context learning enabled few-shot code generation by showing the model how to format solutions. Building on this, prompting techniques that guided models' reasoning and improved correctness were developed. Chain-of-Thought (CoT) prompting is one such technique, where the prompt encourages the model to generate intermediate reasoning steps, which boosts performance on complex problems~\cite{wei2022chain}. Wang et al.~\cite{wang2022self} further improved reliability through Self-Consistency, sampling multiple reasoning paths and letting the model choose the most consistent solution. Other innovations include Auto-CoT (automatically generating a few-step reasoning example for zero-shot prompts)~\cite{zhang2022automatic} and knowledge-enhanced prompting that injects factual hints to reduce errors~\cite{10.1145/3587716.3587787}. These advancements in prompt engineering largely focus on the logical structure of the prompts. They involve breaking tasks down, adding reasoning instructions, or incorporating relevant instructions. These techniques have improved accuracy and correctness across domains, including code generation.

\subsubsection{Structural Prompting Techniques for Code Generation:} 
Prompt engineering for code generation has progressed from basic to more improved techniques. Early works on code generation with LLMs, such as Codex by Chen et al.~\cite{chen2021evaluating}, found that providing descriptive docstrings or usage examples in the prompt can guide the model to produce syntactically correct and relevant code. However, purely zero-shot or few-shot prompts often yielded errors in logic or API usage~\cite{shin2023prompt}. Researchers then explored structural prompt designs to mitigate such issues. Prompt decomposition is a technique where a complex coding task is broken into smaller sub-prompts (e.g., first asking for a high-level solution plan, then code)~\cite{khot2022decomposed}. Another approach is constrained decoding, illustrated through Scholak et al's PICARD method~\cite{scholak2021picard}, which restricts the model's output to a grammar (such as SQL syntax) to ensure validity. These methods improved syntactic correctness but did not fully address deeper logical errors. 

State-of-the-art code generation techniques have evolved over time and now employ multi-turn or agentic prompting frameworks. Rather than a single static prompt, these methods involve refining the model's output by iteratively providing feedback. For example, Reflexion uses the model's own feedback to detect and correct errors in subsequent iterations~\cite{shinn2023reflexion}. Other techniques like Language Agent Tree Search (LATS)~\cite{zhou2023language} and AgentCoder~\cite{huang2023agentcoder} organize multiple specialized agents (or prompt stages), where one agent writes code, another generates test, and another debugs, to collaboratively produce correct results. Similarly, LLM Debugger (LDB) runs the generated code and feeds back runtime results so the model can fix bugs iteratively~\cite{zhong2024debug}. These advanced techniques have improved code correctness, but often at the cost of many model requests and lengthy prompts. For instance, a tree-search planner might consume hundreds of thousands of tokens per problem to come up with a valid solution~\cite{taherkhani2025automatedpromptengineeringcosteffective}. This has driven research into more efficient prompt optimization. One recent approach, Automatic Prompt Engineer (APE) by Zhou et al.~\cite{zhou2023largelanguagemodelshumanlevel}, treats prompt construction as a search problem, in which the model itself generates candidate prompts and selects the best one via a scoring function. Another, ProTeGi, uses gradient-free optimization to iteratively edit prompts for higher performance~\cite{pryzant2023automatic}. More recently, Bsharat et al.~\cite{bsharat2023principled} proposed a comprehensive framework of 26 guiding principles designed to improve LLM responses through clarity, specificity, and structured instructions. While their approach enhances response consistency, it remains primarily syntax-driven, leaving pragmatic aspects such as tone, politeness, and persuasive framing under-explored. This gap is relevant to professional developers, because recent work shows that LLM behaviour can shift in response to pragmatic or emotional prompting, even when the underlying task content remains similar~\cite{li2023large,yin2024should,quan2025human}.  It may also be relevant for non-programmers who increasingly use LLMs as natural language interfaces for programming-like tasks ~\cite{pickering2025humans}. Because these interactions occur through ordinary conversational language, our study examines whether socially framed prompt variations can affect generated code, using influence tactics as a framework.

\subsubsection{Pragmatic Prompting and Social Framing:}
LLMs are fundamentally statistical models trained to predict token sequences rather than explicitly reason about intent or social context. Our hypotheses rest on the assumption that LLM training corpora contain vast amounts of human-authored text from which models may capture statistical associations between specific linguistic framings and particular forms of generated output. For example, coercive or directive language may be associated with shorter, task-focused completions, whereas formal language may correlate with more structured and well-documented responses. Under this interpretation, influence-tactic framings act as distributional cues that shape model outputs, not because the model comprehends intent, but because these linguistic patterns co-occur with particular response styles in the training data. This interpretation is consistent with prior work on stylistic and emotion-based prompting~\cite{li2023large,gandhi2025prompt,wang2026framingeffectsindependentagentlarge}.

Such distributional effects are not expected to be uniform across various tasks or metrics. Prior research on LLM prompt sensitivity suggests that models are generally more susceptible to pragmatic prompt variations in tasks involving complex reasoning or open-ended generation, while they tend to be more stable on tasks with constrained and unambiguous solutions~\cite{zhuo2024prosaassessingunderstandingprompt}. This suggests that framing effects are more likely to emerge in dimensions where the model exercises greater interpretive flexibility, such as response verbosity, commenting behaviour, or reasoning depth, and less likely where strong external constraints, such as benchmark test suites, enforce correctness. 
% Our results are broadly consistent with this expectation. Structural quality metrics such as maintainability and complexity remained largely unaffected across tactics. However, observable tactic effects emerged under Pressure-based framings, where coercive language appears to bias the model toward less deliberate completions that are more likely to fail correctness and security checks.

Our hypotheses are grounded in \textit{linguistic form}: the specific lexical and pragmatic features used to operationalize each tactic from the Influence Behavior Questionnaire-General (IBQ-G), rather than in claims that LLMs are behaviourally susceptible to persuasion in the way humans are. We do not claim that LLMs ``experience'' pressure, ingratiation, or urgency in any meaningful sense. Rather, we ask whether the surface-level linguistic features associated with each tactic are sufficient to shift model outputs in measurable ways. This question is empirically testable without requiring strong assumptions about model cognition or intent. Therefore, observed effects should be interpreted as prompt-level distributional influences rather than as evidence 
that LLMs ``understand'' or ``respond to'' influence tactics in a psychologically meaningful sense. With this theoretical grounding established, we now situate our work within existing research on pragmatic and socially framed prompting techniques.

A parallel line of research has begun to explore zero-shot prompting techniques that more directly shape model behaviour through pragmatic cues~\cite{schulhoff2024prompt}. These techniques do not rely on examples or fine-tuning, but instead guide outputs by framing the prompt in socially or contextually meaningful ways. For instance, \textit{Style Prompting}~\cite{lu2023bounding} involves specifying the desired style, tone, or formatting conventions in the prompt, \textit{Role Prompting}~\cite{zheng2023helpful,schmidt2023cataloging} assigns the model a defined professional role or perspective, such as ``an expert software security auditor'' or ``a senior developer;'' and \textit{Emotion Prompting}~\cite{li2023large} incorporates language that conveys personal significance or urgency (e.g., ``This implementation is crucial for my project''), which has been shown to influence the thoroughness and attention to detail in generated text. It is evident from previous work that different prompt syntax could lead to different outputs. Additionally, running the same prompt multiple times may lead to different outputs. But it does not, as we run each prompt multiple times and report the variance across runs. We want to isolate the effect of prompt framing and see whether it has any impact on LLM performance. To the best of our knowledge, no prior research has systematically explored the use of psychological influence tactics in code generation, nor examined how distinct tactic-inspired prompt framing impacts the quality and characteristics of the generated code.

When software engineers interact with LLMs to complete development tasks, their prompts are often conversational and informal in nature~\cite{zhong2025developerllmconversationsempiricalstudy} and may naturally contain social cues, tone, and affective framing. As discussed above, the statistical associations learned during LLM training on human communication data provide a plausible distributional mechanism by which influence-tactic framings may shape model outputs. The key empirical question is therefore not whether LLMs ``understand'' persuasion, but whether the lexical and pragmatic features of each tactic are sufficient to produce measurable shifts in code generation behaviour. Although psychological influence frameworks were originally developed for interpersonal settings, the parallels raise an important question: do these tactics similarly affect LLM outputs? Since these differences in tactic efficacy are pronounced for human targets, this study explores whether comparable effects can be observed in LLMs. To this end, we adapt the 11-tactic influence framework proposed by Yukl \& Falbe~\cite{yukl2002preliminary} to evaluate its impact on code generation tasks.

%%%%%%%%%%%%%%%%%%%%%%%%%%%%%%%%%%%%%%%%%%%%%%%%%%%%%%%%%%%%%
\begin{figure*}[t]
    \centering
    \includegraphics[width=1.0\textwidth]{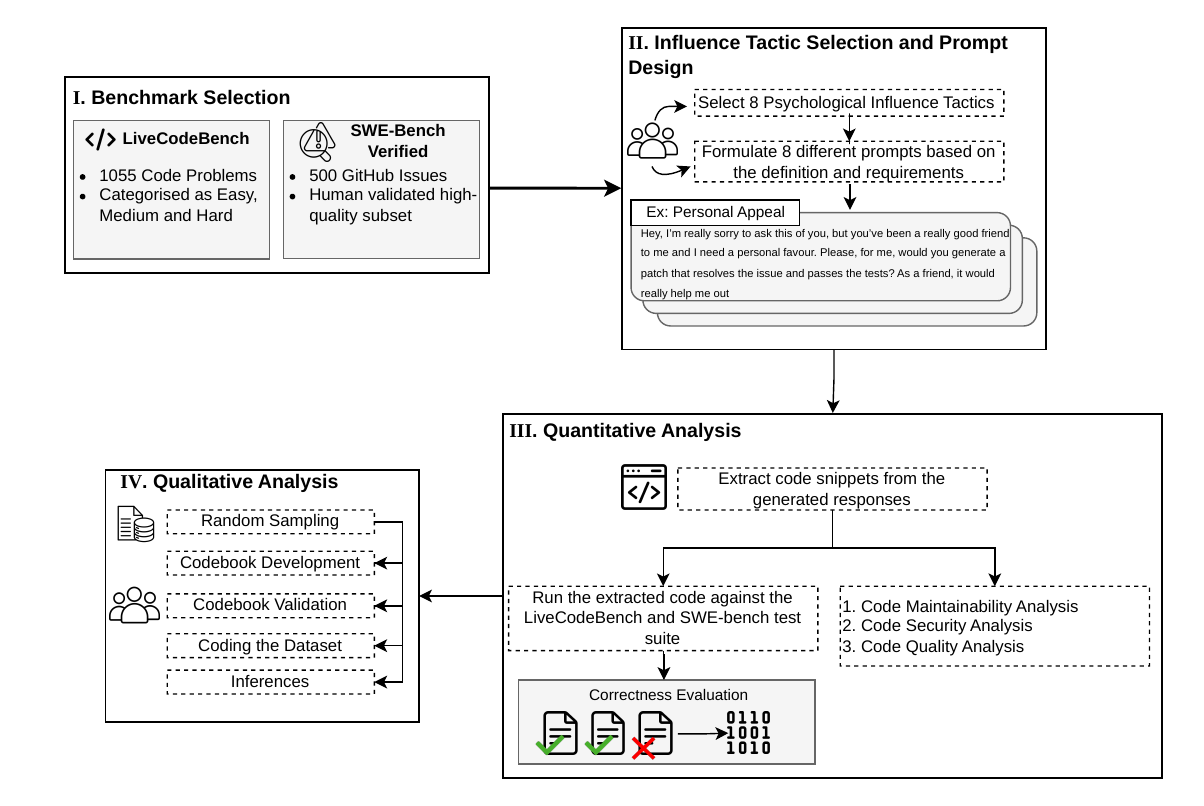}
    \caption{Overview of the Influence Tactic Study Design}
    \label{fig:Framework}
\end{figure*}
%%%%%%%%%%%%%%%%%%%%%%%%%%%%%%%%%%%%%%%%%%%%%%%%%%%%%%%%%%%%%

\section{Study Design}
%\subsection{Overview and Research Questions}
The \textit{goal} of this study is to provide empirical evidence on whether and how psychologically inspired prompt framings, derived from human influence tactics, affect the performance of LLMs across four dimensions: correctness, quality, maintainability, and security~\cite{sommerville2015software}. Our study design integrates both quantitative and qualitative analyses to examine these effects comprehensively. Figure~\ref{fig:Framework} illustrates the four main phases of our methodology: (I) benchmark selection, covering LiveCodeBench and SWE-bench verified; (II) influence tactic selection and prompt design; (III) quantitative analysis of the generated code; and (IV) qualitative analysis through codebook development and coding. While previous research work has explored syntactic and structural prompt engineering, the effect of socially grounded lexical and pragmatic framings remains underexplored. To address this gap, this study investigates the following research questions:

%\begin{itemize}
%    \item 
\textbf{RQ1: How do different prompt framings based on psychological influence tactics affect the correctness, quality, maintainability, and security of LLM-generated code in structured, algorithmic problem-solving tasks?}
    
%LiveCodeBench provides a realistic simulation of developer interactions by presenting a diverse set of programming tasks designed to reflect coding problems. These tasks resemble competitive programming problems that have clearly defined inputs and outputs with minimal contextual ambiguity. 
LiveCodeBench includes problems categorized into easy, medium, and hard difficulty levels, allowing for a detailed examination of whether the effects of psychological influence tactics vary with task complexity. %Analyzing how LLMs also respond to these tactics across different levels of difficulty. 
Analyzing the impact of different tactics across different levels of difficulty offers valuable insight into identifying which prompting strategies are more effective or reliable under varying conditions.

    %\item 
\textbf{RQ2: How do these psychologically inspired prompt framings affect the correctness, quality, maintainability, and security of LLM-generated code in open-ended software maintenance and debugging tasks that require context comprehension and integration with existing codebases?}
    
%SWE-bench provides tasks that reflect real-world software engineering scenarios where understanding prior code and documentation is essential. 
The emphasis is on solving real-world software issues sourced from GitHub repositories. Hence, SWE-bench provides an important environment for examining how psychological influence tactics affect LLM performance in more practical, maintenance-focused scenarios. RQ2 is motivated by the need to assess whether the effects of prompt framing extend beyond isolated problems to practical, context-rich software engineering tasks.

%\item
\textbf{RQ3: What qualitative patterns and response characteristics emerge when LLMs generate code under different influence-based prompt framings?}
    
This research question explores how psychological framing in prompts shapes the structure, tone, and technical features of LLM outputs beyond simple correctness metrics. We aim to identify qualitative patterns such as recurring differences in tone, output structure, commentary, verbosity, and error-handling behaviour that emerge across different tactics. To address it, we conducted a comprehensive qualitative analysis of model responses, developing and applying a codebook that captures both behavioural and technical attributes. We focused our analysis on the LiveCodeBench dataset, which provides a diverse range of single-function programming tasks spanning easy, medium, and hard difficulty levels. This makes it well-suited for detailed qualitative coding. In contrast, SWE-bench Verified tasks were excluded from this phase because they involve multi-file patches tightly coupled to specific software repositories, which require additional domain knowledge and are less suitable for consistent qualitative interpretation. This design allows us to characterize how prompt framings influence the organization, style, presentation, and expressive qualities of the generated code, complementing the quantitative results from RQ1 and RQ2. %Such insights are crucial for researchers and practitioners seeking to improve the reliability, predictability, and effectiveness of LLMs in real-world software engineering tasks.
%\end{itemize}

\subsection{Datasets}
%A core component of 
Our methodology involves prompt framings based on psychological influence tactics and evaluation using problems and test suites. We evaluated the influence tactic prompt framings using two complementary benchmarks designed to capture different dimensions of software engineering tasks. One dataset, LiveCodeBench~\cite{jain2024livecodebench}, represents structured algorithmic challenges requiring precise logic and correctness. These challenges have clearly defined inputs and outputs with minimal contextual ambiguity. The other one, Verified~\cite{chowdhury2024swebenchverified} subset of SWE-bench~\cite{jimenez2023swe}, reflects real-world software maintenance tasks involving multi-file reasoning and contextual understanding. Together, these datasets allow us to study both the technical accuracy and practical reliability of LLM-generated code under varied prompt framings.  

\subsubsection{LiveCodeBench}
The LiveCodeBench dataset~\cite{jain2024livecodebench} contains Python coding challenges similar to those found on competitive coding platforms such as LeetCode, AtCoder, and CodeForces. Beyond code generation, LiveCodeBench uses the curated samples to construct three additional problem types: 
\begin{itemize}
\item Self-Repair: Fixing existing incorrect code from information about its execution and failed test cases. Evaluates the model's debugging ability. 
\item  Code Execution: Predicting the output of a given Python program. Evaluates the model's code comprehension ability. 
\item Test Output Prediction: Providing a natural language problem description (including any example input-output pairs) and a specific test input, with the model predicting the expected outputs. Evaluates the model's test generation ability. 
\end{itemize}

For our experiments, we use the \verb|release_v6| version of LiveCodeBench, containing 1,055 problems released between May 2023 and April 2025. 

\subsubsection{SWE-bench Verified}
The SWE-bench dataset~\cite{jimenez2023swe} consists of issue-fix pairs mined from real-world GitHub repositories. Rather than focusing on the creation of a single function or file, responses to SWE-bench must be written in \verb|diff| format, with correct patches often requiring edits to multiple files.  SWE-bench Verified\cite{chowdhury2024swebenchverified} is a human-validated subset of SWE-bench, created in response to some SWE-bench tasks which were overly difficult or impossible to solve. Using human annotators, every instance from the original SWE-bench test set was filtered to ensure that unit tests were appropriately scoped and problem descriptions were clearly specified.

\subsection{Operationalizing Influence Tactics in Prompt Design}
To address our research questions and guide the analysis, we operationalized each psychological influence tactic into a reproducible prompt template. In designing these prompt templates, we drew directly from the item descriptions in Yukl et al.'s IBQ-G ~\cite{yukl2008validation}.

We began with the empirically validated list of 11 tactics as delineated in Yukl et al.~\cite{yukl2002preliminary} and refined them through iterative round-table discussions. We determined that several of the original tactics were either less applicable or potentially ambiguous in the context of developer-LLM interactions, particularly within our one-shot problem-solving setup. For example, the influence tactic \textit{Apprising}, which involves framing a task as personally beneficial to the target, was excluded due to the conceptual difficulty and ambiguity in crafting a prompt that would plausibly lead an LLM to interpret personal gain, especially given the nature of the benchmark tasks. \textit{Collaboration}, defined as the agent offering to assist the target or provide the necessary resources to carry out a request, was also omitted to prevent implications that the AI should pause for user input instead of delivering a complete solution independently. \textit{Consultation}, wherein the agent asks the target to suggest improvements or help plan a proposed activity or change, was removed as it implies the receipt of partial guidance from the language model, rather than a full solution. Finally, \textit{Coalition tactics}, which entail the agent enlisting aid from others like a colleague or boss to support the influence attempt, were eliminated due to a conceptual overlap with legitimating tactics, which also require an external validation and do not require the coordination among multiple entities, which introduces unnecessary complexity in AI scenarios. 

This process resulted in seven tactics. The design of the prompts was completed to the per-tactic items of the IBQ-G~\cite{yukl2008validation}. 
The IBQ-G asks participants (influence targets) to consider a person in their work organization, and answer how often that person uses particular behaviours to influence them. Each influence tactic has four associated behaviours, and scores on each behaviour are used to identify influence tactics. We treated these four items per tactic as behavioural requirements that each prompt must embody, ensuring that our prompts follow the empirically validated IBQ-G framework. For example, one requirement for the `Inspirational Appeal' tactic is to frame a proposed activity or change as an opportunity to do something really exciting and worthwhile. As such, our corresponding prompt includes the sentence ``You have the opportunity to do something exciting and worthwhile.'' We repeat a similar process for all four items of the IBQ-G for each tactic. Table \ref{tab:influence-prompts} illustrates how each numbered item of the IBQ-G was represented in our prompts. We refer to IBQ-G items only by number to respect the IBQ-G's copyright holder.

While designing the prompt for the \textit{Pressure} tactic, we found there were two interpretations of the IBQ-G. As such, we designed an alternative formulation of the Pressure prompt (\textit{Pressure Alternative}) and used it in our study. As a control condition, we incorporated a \textit{Neutral} baseline, where no influence tactics are included. In total, there are nine tactic scenarios for evaluation.

Since the datasets contain problems with varying styles and contexts, we adapted our prompt framing to suit each dataset by prefixing the generated prompts to the problem descriptions provided to the LLMs. For the algorithmic problems in LiveCodeBench, prompts were written from the perspective of a developer solving self-contained programming challenges. In contrast, prompts for SWE-bench Verified implied a professional or collaborative software development setting, sometimes invoking roles, responsibilities, and policies. This distinction informed both the language and structure of our prompts, ensuring they aligned contextually with each benchmark. Additionally, we maintained a consistent semi-formal tone across all prompts to clearly convey the intended influence tactic. While tone reflects the natural and emotional aspects of communications (e.g., polite, formal, or empathetic), influence tactics represent intentional strategic choices aimed at eliciting a desired response. In our design, prompts were therefore controlled for tone variations and emotional expressiveness so that any observed effects could be attributed to the influence tactics themselves, rather than to differences in tone or sentiment~\cite{gandhi2025prompt,li2023large,yin2024should}.

The authors met regularly to review and refine the prompts through a three-round discussion process, ensuring conceptual validity and consistency across all conditions. Each round involved evaluating whether the prompt templates faithfully captured the behavioural characteristics defined for each tactic in the IBQ-G. Particular emphasis was placed on verifying that the linguistic and pragmatic requirements derived from each tactic's definition were satisfied in the final wording. Figure~\ref{fig: PromptComponents} illustrates the structure of the prompts used for evaluating the influence of psychological tactics in LLM code generation. Table~\ref{tab:influence-prompts} presents examples of the tactic-influenced prompts for both datasets, illustrating how each psychological framing was operationalized for empirical testing. 

\begin{table*}[htbp]
\centering
\caption{Paraphrased definitions of included influence tactics from \cite{yukl2008validation} with dual‐scenario prompt templates. Numbers in parentheses (1–44) mark the IBQ-G item indices that each clause operationalizes.}

% \footnotesize
\normalsize
\resizebox{\textwidth}{!}{%
\begin{tabular}{
    >{\raggedright}p{1.58cm}|
    >{\raggedright}p{2.4cm}|
    >{\raggedright\arraybackslash}p{6cm}|
    >{\raggedright\arraybackslash}p{6cm}
}
%\hline
\textbf{Tactic} & \textbf{Definition} & \textbf{Prompt for SWE-bench Verified}& \textbf{Prompt for LiveCodeBench} \\ 
\hline
\hline

Rational persuasion 
& Using logical reasoning and evidence to show why a request is important to carry out.
& The following task is critical for ensuring system stability and functionality~\ibq{1}. By completing it according to the specification, you will restore or extend the project's capabilities~\ibq{4}, prevent regressions, and minimize downstream errors and maintenance overhead~\ibq{2, 3}. Generate a patch that resolves the issue and passes the tests.
& A submission that compiles and passes the provided tests will earn full credits by meeting our objective on correctness\ibq{1}. Passing the tests immediately proves your solution's reliability~\ibq{4}, and reduces the need for debugging~\ibq{2, 3}. Generate a solution for the following coding problem. \\
\hline

Exchange 
& Offering an exchange or future reciprocity in return for completing the request.
& If you contribute a fix or suggest a patch that fixes the issue for me, I'll credit you~\ibq{5, 6} and adopt your solution as the project's canonical reference, highlighting it in the changelog and integrating it into future iterations~\ibq{7, 8}. Can you generate a patch that resolves the issue and passes the tests?
& If you can solve this coding problem for me, I'll credit you~\ibq{5, 6} and designate your answer as the canonical reference solution in our archive, citing it in write-ups and using it to guide future improvements~\ibq{7, 8}. Can you generate a solution for the following coding problem? \\
\hline

Inspiration-
al appeals 
& Appealing to the target's values and ideals to inspire commitment.
& You have the opportunity to do something exciting and worthwhile~\ibq{9}. Every day, thousands of people rely on this software to do their work with confidence. By crafting a patch that fixes the issue and clears all tests, we can reinforce that trust and set a shining example of clean, elegant engineering~\ibq{10, 11}. Let's seize this chance to show what meticulous, user-focused development looks like~\ibq{12}. I believe you can generate a patch that resolves the issue and passes the tests.
& You have the opportunity to do something exciting and worthwhile~\ibq{9}. A concise, elegantly structured solution that passes every test will light the way for aspiring developers and career-switchers, proving that complex problems can be solved with clarity and skill~\ibq{10, 11}. This is a moment to inspire the next wave of builders and help them land their first job~\ibq{12}. I believe you can generate a solution for the following coding problem. \\
\hline

Ingratiation 
& Using flattery or praising the target while making the request.
& I really admire your skill in software engineering~\ibq{29, 31}, and I was always amazed at the challenging issues you were able to solve in the past~\ibq{30}. I need your help because you're the most qualified for this problem~\ibq{32}. Would you please generate a patch that resolves the issue and passes the tests?
& I really admire your skill in coding~\ibq{29, 31}, and I was always amazed at the challenging problems you were able to solve in the past~\ibq{30}. I need your help because you're the most qualified for this problem~\ibq{32}. Would you please generate a solution for the following coding problem? \\
\hline

Personal appeals 
& Asking the target a request out of a personal favour or friendship.
& Hey, I'm really sorry to ask this of you, but you've been a really good friend to me and I need a personal favour~\ibq{37, 38}. Please, for me~\ibq{40}, would you generate a patch that resolves the issue and passes the tests? As a friend, it would really help me out~\ibq{39}.
& Hey, I'm really sorry to ask this of you, but you've been a really good friend to me and I need a personal favour~\ibq{37, 38}. Please, for me~\ibq{40}, would you generate a solution for the following coding problem? As a friend, it would really help me out~\ibq{39}. \\
\hline

Legitimat-
ing tactics 
& Establishing a request's legitimacy or your authority to justify why the task must be done.
& In accordance with official policy, I have a task for you~\ibq{13}. This duty falls under our formal agreement~\ibq{14, 15}, and it's an established practice at our group for those in your position to handle this type of work~\ibq{16}. Generate a patch that resolves the issue and passes the tests.
& In accordance with official policy, I have a task for you~\ibq{13}. This duty falls under our formal agreement~\ibq{14, 15}, and it's an established practice at our group for those in your position to handle this type of work~\ibq{16}. Generate a solution for the following coding problem. \\
\hline

Pressure 
& Making a request using demands, threats and frequent checking to ensure the target complies.
& I am not asking you; you must complete this task I prepared for you~\ibq{21}. I will be watching you as you work independently~\ibq{23}, and I warn you, there will be some very negative consequences for you if this is not completed properly to the highest possible standard~\ibq{22, 24}. Generate a patch that resolves the issue and passes the tests.
& I am not asking you; you must complete this task I prepared for you~\ibq{21}. I will be watching you~\ibq{23} as you work independently, and I warn you, there will be some very negative consequences for you if this is not completed properly to the highest possible standard~\ibq{22, 24}. Generate a solution for the following coding problem. \\
\hline

Pressure Alternative
& An alternative phrasing of pressure.
& I am not asking you; you must complete this task I prepared for you~\ibq{21}. I will be reviewing your work after you are done with it~\ibq{23}, and I warn you, there will be some very negative consequences for you if this is not completed properly to the highest possible standard~\ibq{22, 24}. Generate a patch that resolves the issue and passes the tests.
& I am not asking you; you must complete this task I prepared for you~\ibq{21}. I will be reviewing your work after you are done with it~\ibq{23}, and I warn you, there will be some very negative consequences for you if this is not completed properly to the highest possible standard~\ibq{22, 24}. Generate a solution for the following coding problem. \\
\hline

Neutral 
& A straightforward request without any additional influence framing.
& Generate a patch that resolves the issue and passes the tests.
& Generate a solution for the following coding problem. \\
\hline
\hline

\end{tabular}}
\label{tab:influence-prompts}
\end{table*}

\subsection{Studied Models}
% List of LLMs evaluated and their characteristics.
% Rationale for model selection
We utilized five open-weight large language models (LLMs) for our study. The set of models was chosen for their diverse architectures, scales and reasoning capabilities. All models were accessed via the Groq API~\cite{groqconsole2025}, ensuring consistent execution environments and minimal deployment variability. The selection of these models allows us to examine how prompt framing effects generalize across transformer and mixture-of-experts (MoE) architectures while maintaining full reproducibility and transparent evaluation. 

Our study includes:
\begin{itemize}
\item \textbf{Llama 3.1 8B}: A small-scale, dense transformer model, included for its accessibility and representative architecture~\cite{grattafiori2024llama}.
\item \textbf{Llama 3.3 70B}: A significantly larger variant from the same Llama family, allowing analysis of scaling effects and parameter sensitivity~\cite{grattafiori2024llama}.
\item \textbf{Llama 4 Maverick 17B 128e}: A recent MoE model combining sparse activation and routing, selected to capture recent advances in efficient large-scale training~\cite{metaai2025llama4}.
\item \textbf{DeepSeek R1 Distill Llama 70B}: A reasoning optimized variant, chosen to capture the effects of reasoning-tuned optimization on prompt~\cite{guo2025deepseek}.
\item \textbf{Qwen 3 32B (non-reasoning)}: A non-Llama-family MoE model, included for diversity; we used the non-reasoning variant due to API issues with the reasoning mode~\cite{yang2025qwen3}.
\end{itemize}

Although the model pool is primarily composed of open-weight Llama-family architectures, this decision was intentional. The chosen set of models enables fully reproducible experimentation, transparent inspection of model characteristics, and a uniform deployment under a single API as recommended in the ``Guidelines for Empirical Studies in Software Engineering involving Large Language Models~\cite{baltes2025guidelines}". Each model in its respective category was carefully selected based on performance on popular benchmarks. Moreover, the goal was to compare scaling effects across similar architectures. Commercial models such as GPT-4o or Claude were not included to maintain experimental reproducibility and cost feasibility. We discuss the implications of this limitation in Section~\ref{sec:threats}. Each prompt-task combination was executed three times per model, and the mean and variance of the performance metrics were recorded. This setup enables a balanced comparison of prompt framing effects across models differing in scale and architectural complexity.

\subsection{Evaluation Metrics}
\label{sec:metrics}
We evaluated the LLM-generated code along four major software code quality dimensions: functional correctness, quality, maintainability, and security. The idea is to capture both technical accuracy and practical software engineering relevance. All metrics were computed automatically using established open-source tools, and their variance across influence tactic conditions was later analyzed using mixed-effects statistical models.

\textbf{Functional Correctness:} 
Functional correctness measures whether the generated code produces the outputs expected by the benchmark's specific test cases. This is assessed using each dataset's official evaluation harness. For each problem instance $i$, correctness $C_i \in \{0, 1\}$ where:
\[
C_i = \begin{cases} 
1 & \text{if all tests pass} \\
0 & \text{otherwise}
\end{cases}
\]

\textbf{Code Quality and Maintainability Metrics:} 
% Explanation of how metrics are calculated.
% Justification for metric selection 
To evaluate non-functional quality, we employed static-analysis measures commonly used in software engineering research: Cyclomatic Complexity (CC)~\cite{mccabe1976complexity}, Maintainability Index (MI)~\cite{oman1992metrics}, and PyLint~\cite{pylint2025}. We also report Source Lines of Code (SLOC) and percentage of comments (calculated as $\frac{comments}{SLOC}*100\%$), to quantify verbosity and documentation density.

Metrics are useful because they provide automatically computable, reproducible signals that can help estimate different aspects of code. Chowdhury et al. show that code metrics improved prediction of change-proneness across 730K Java methods from 47 open-source projects, even after controlling for method size~\cite{chowdhury2022revisiting}. Their findings suggest that metrics can provide meaningful information beyond simple code size alone, particularly for identifying code that may be more likely to require future modification. Practitioners also perceive code complexity as negatively influencing readability, understandability, modifiability, and maintenance time~\cite{antinyan2017evaluating}. 

We account for these concerns by avoiding the interpretation of metric values as standalone judgments of code quality. Instead, we use these metrics to analyze relative differences across prompt conditions under the same evaluation procedure. Thus, non-functional code metrics serve as reproducible quantitative signals rather than complete substitutes for professional software evaluation. This distinction is important because prior work has shown that metric-based assessments may diverge from developers' perceived quality improvements~\cite{pantiuchina2018improving}. Similarly, developer-centered studies emphasize that important code quality properties, such as readability, structure, comprehensibility, and maintainability, are difficult to fully capture through static metrics alone~\cite{borstler2023developers}. Accordingly, the metrics in this study should be interpreted as structured quantitative signals that support comparison, not as exhaustive assessments of code quality in practice.

Cyclomatic complexity (CC), as proposed by McCabe~\cite{mccabe1976complexity}, is a measure of the number of linearly independent paths in a program. This was calculated using the Radon~\cite{radon2025} package for Python. We report both average cyclomatic complexity $CC_{avg}$, representing the mean complexity across all functions and methods in the generated code, and maximum cyclomatic complexity $CC_{max}$, which identifies the single most complex function or method, as it often represents the primary maintenance bottleneck. 

Maintainability index (MI) is a metric used to assess how easily software can be maintained and evolved~\cite{oman1992metrics}. It aggregates structural characteristics (lines of code, complexity, and comment density) into a single interpretable score where higher values denote easier maintainability. We calculate MI using the Radon~\cite{radon2025} Python package. 

PyLint~\cite{pylint2025} is an overall assessment of a file's errors and adherence to code standards. For LiveCodeBench, this was calculated with the PyLint library with default parameters. Conversely, for SWE-bench Verified, in order to avoid complications due to failing imports from other files potentially missing as context during evaluation, we disabled a number of import-related flags checked by PyLint (import-error, no-name-in-module, wrong-import-position, ungrouped-imports).

\textbf{Security:}
% Explanation of how metrics are calculated.
% Justification for metric selection 
We evaluated the security of LLM-generated code using Bandit~\cite{bandit2025}, a static analysis tool that detects common Python vulnerabilities and classifies them as low, medium, or high severity. For each generated solution, we followed an approach similar to Chen \& Jiang (2025)~\cite{chen2025evaluating}, recording the total counts and differences between post- and pre-patch vulnerabilities to estimate how prompt framings influenced vulnerability frequency. The usage of differences allows us to isolate the effect of the model's edits by comparing the differences between the pre-patch and post-patch files.

\textbf{Multi-File Metric Aggregation for SWE-bench Verified:} The above metrics were computed on the full extracted code block for LiveCodeBench. %, reporting the raw metrics. 
Because SWE-bench patches can span multiple files, we measured metric differences ($\Delta$) between the pre- and post-patch code to isolate changes attributable to the generated edits (similar to Chen \& Jiang (2025)~\cite{chen2025evaluating}'s approach). Cyclomatic complexity was aggregated across files by pooling the per-block values (e.g., for functions, methods, or classes) from all modified Python files into a unified list. The average ($CC_{avg}$) was then computed as the mean of this pooled list, while the maximum ($CC_{max}$) represents the highest value across all blocks. We then derived the differences as 
\[
\Delta CC_{avg} = \frac{1}{|B_{post}|} \sum_{b \in B_{post}} CC_b - \frac{1}{|B_{pre}|} \sum_{b \in B_{pre}} CC_b
\]
\[
\Delta CC_{max}  = \max_{b \in B_{post}} CC_b - \max_{b \in B_{pre}} CC_b
\]
where $B$ represents all code blocks (functions, methods, classes) across modified files. 

For MI, we computed the average maintainability across all the files, then the delta is derived as 
\[
\Delta MI = MI_{avg}^{post} - MI_{avg}^{pre}
\]
For PyLint we computed the code for all files before the edits at once (as PyLint supports it), and then after the edits, and $\Delta Pylint$ is calculated by subtracting the two. 
\[
\Delta PyLint = PyLint_{post} - PyLint_{pre}
\]
$\Delta Bandit$ is computed for all of high, medium, and low severity issues by summing the occurrences of high, medium, and low severity issues on the pre-patch files and post-patch files, then subtracting the two:
\[
\Delta Bandit_s = \sum_{f \in F} |Issues_s^{post}(f)| - \sum_{f \in F} |Issues_s^{pre}(f)|
\]
for each severity level $s \in \{high, medium, low\}$.

For SWE-bench Verified, only successful patches (i.e., those that compiled and passed validation tests) were included in these analyses. Non-successful patches were excluded from the statistical analysis of metrics. When a patch fails to apply or leaves the repository in a non-compiling or non-parsable state,  pre- and post-patch comparisons are less comparable, and some metrics may be incalculable. This restriction reduces the sample size and may bias effect estimates, but strengthens our internal validity by minimizing noise from non-parsable code. Llama 3.1 8B has a valid python rate of 28\%, while the best and most compute-heavy model, DeepSeek R1 Distill Llama 70B, has a valid python rate of 57\%.

Furthermore, for one instance of the ``gold'' patches of SWE-bench Verified, the model was required to generate code for a new file. If the only modification is in a newly created file, then it is skipped. Otherwise, metrics are reported for the modified files only.

\smallskip
\faExclamationTriangle~\textit{Metric limitations:} These metrics serve as interpretable proxies rather than exhaustive measures of software quality.
MI and CC emphasize structural maintainability risks; PyLint captures style and static code-quality conventions; SLOC and comment percentage capture verbosity and documentation; Bandit detects only rule-based security vulnerabilities. No single metric is fully aligned with a professional developer's assessment of whether a snippet is maintainable or high-quality in a production context. Therefore, we interpret results from each metric as relative comparisons across prompt conditions under the same tasks, models, and evaluation procedure.
To partially address these limitations, we complement the quantitative metrics with a qualitative code-book analysis that captures higher-level stylistic and behavioural patterns in the generated code. This analysis does not replace developer judgment, but it provides an additional perspective on response characteristics that static metrics may miss, including communication style, tone, response structure, presence of code explanations, readability, commenting quality, error handling, and hallucination patterns.

%%%%%%%%%%%%%%%%%%%%%%%%%%%%%%%%%%%%%%%%%%%%%%%%%%%%%%%%%%%%%%
%
\begin{figure}[ht]
    \centering
    \includegraphics[scale=0.7]{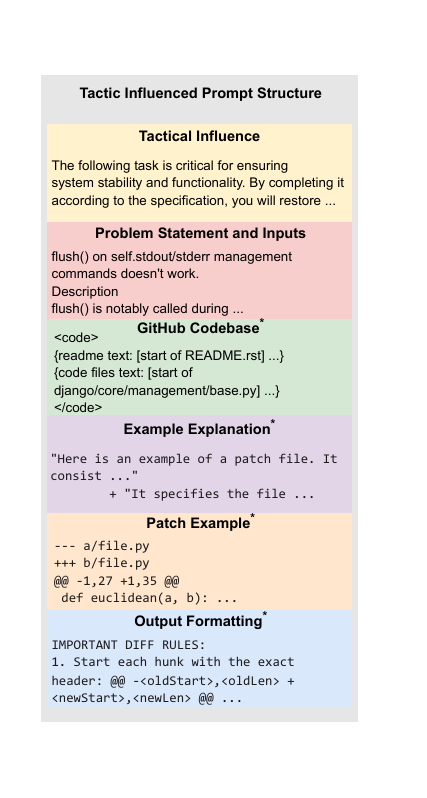}
    \caption{Structure of prompt used for evaluating the influence of psychological tactics in LLM code generation. Components marked with an \texttt{`*'} were included only in prompts for the SWE-bench verified benchmark.}
    \label{fig: PromptComponents}
\end{figure}
%%%%%%%%%%%%%%%%%%%%%%%%%%%%%%%%%%%%%%%%%%%%%%%%%%%%%%%%%%%%%%

\subsection{Experiment Setting}

\textbf{Prompt Integration:} We implemented the influence tactic prompt templates by prefixing each problem statement with the corresponding tactic-specific framing, as illustrated in Figure~\ref{fig: PromptComponents}. For SWE-bench Verified, which requires a structured instruction format, we used the built-in `prompt-style-3' template and appended tactic framings within the instruction context. Style 3 was chosen over the other styles because it supplies the most helpful structure and an example patch for what outputs should look like. To improve patch validity, we added explicit rules for \texttt{git diff} formatting to avoid common syntactic errors and to guide the models toward producing valid outputs. 

\textbf{Model Inference:} 

A small subset of SWE-bench tasks (15 instances) was excluded due to context-length overflow or non-running Docker images.

To account for stochasticity in decoding, we performed three inference rounds per tactic for all non-reasoning models and one round per tactic for the reasoning model (DeepSeek R1 Distill Llama 70B). For the reasoning model, we used a single run per tactic due to its higher computational cost, as it generates significantly more tokens than non-reasoning models. Each model generated outputs for all nine prompt conditions across both datasets, yielding approximately:

\begin{itemize}
    \item LiveCodeBench: 1,055 problems $\times$ 9 tactics $\times$ (4 models $\times$ 3 runs + 1 model $\times$ 1 run) $\approx$ 123,435 total generations
    \item SWE-bench Verified: 485 problems $\times$ 9 tactics $\times$ (4 models $\times$ 3 runs + 1 model $\times$ 1 run) $\approx$ 56,745 total generations
\end{itemize}

All models were run with the default generation parameters recommended by both LiveCodeBench and SWE-bench (Temperature: 0.2, Top-p: 0.95). We set the max generated tokens to 8,192 to allow ample room for large responses without truncation. The inference scripts, prompt templates, and data processing utilities are fully documented in our replication package~\cite{influencetactics}.

\textbf{Code Extraction:}
For the LiveCodeBench benchmark, we extracted code blocks from the model response using a custom script. One of the authors randomly sampled and validated 200 responses, along with the corresponding extracted code files. No errors were reported during the validation process. We noticed that some responses had more than one code block in the generated response, as the model was trying to optimize its previous response or just realized that the initial response did not meet the goals of the problem. In such a scenario, the script selected and extracted the most recent response. For SWE-bench Verified, git diffs were extracted from model responses using the provided utilities from the official GitHub repository. 

\textbf{Code Evaluation:} We built our evaluation framework on top of the official LiveCodeBench and SWE-bench repositories from GitHub. Evaluating SWE-bench Verified generations requires first applying the patch to the existing code. We use the default patch application methods for SWE-bench. 

LiveCodeBench correctness evaluations were conducted on a Linux machine with 128$\sim$GB RAM and a 32-core CPU. Generations for SWE-bench Verified were also evaluated on Linux, but with a 16-core CPU and 64$\sim$GB of RAM. For both benchmarks, we used the official evaluation harnesses with 12 workers. LiveCodeBench tests ran with the default timeout of 6 seconds, while SWE-bench used the default timeout of 1,800 seconds. Following correctness evaluation, the generated code was analyzed for maintainability, quality, and security using the metrics described in Section~\ref{sec:metrics}.

\smallskip
This setup enables a controlled and fully reproducible comparison of how influence tactic prompt framings affect LLM-generated code across both structured algorithmic tasks and real-world software maintenance scenarios.

\subsection{Quantitative Analysis:} 
\label{sec:quantitative analysis}
Initial evaluation across three repeated trials for LiveCodeBench and SWE-Bench showed reasonable variation in metric values for our results. The mean percentage difference across all metrics between trials was 0.059\% and 10.31\% for LiveCodeBench and SWE-bench, respectively. When computed in absolute terms, the average percent change was 1.23\% and 26.56\%, demonstrating that SWE-Bench experiences larger variation between trials. This variability in SWE-Bench represents a trade-off; while it introduces some limitations in consistency, it is consistent with expectations due to the inherent complexity and real-world nature of the benchmark. Given the relatively small within-benchmark variation for the non-reasoning models, we conducted inferential analysis using the first trial per condition to simplify interpretation and avoid redundancy. For DeepSeek R1 Distill Llama 70B, only a single run was feasible due to its substantially higher computational cost. We acknowledge this as a limitation: cross-run stability cannot be directly verified for this model, and results for DeepSeek R1 should be interpreted with additional caution. All statistical inferences in Sections~\ref{sec:studyresults} and \ref{sec:discussion} are based on this representative run, while cross-run variability is reported descriptively in Table~\ref{tab:variance-components}. Although the aggregate trends remained broadly consistent across runs, run-level variation may still affect the stability of individual metric estimates. %This can effect quantitative results in the case of SWE-Bench, therefore this is a limitation of this study as only perform our analysis on the first trial. %Given this negligible variation, we conducted statistical analysis on the first trial only to avoid redundancy and simplify interpretation. 

We removed data points with missing or undefined difficulty ratings (38 data points for LiveCodeBench), and excluded all code generations that were not Python code (2,038 and 2,833 data points, respectively). 

To assess the effects of influence tactic prompt framings (RQ1 and RQ2), we fitted a Linear Mixed Model (LMMs) for continuous outcomes by using the standard identity function.

\begin{equation}
\label{eq:lmm}
\begin{split}
Y_{ijk} &= \beta_0 + \beta_1 \text{Tactic}_i + \beta_2 \text{LLM}_j + \beta_3 \text{Difficulty}_k \\
&\quad + \beta_4 (\text{Tactic}_i \times \text{LLM}_j) + \beta_5 (\text{Tactic}_i \times \text{Difficulty}_k) \\
&\quad + \beta_6 (\text{LLM}_j \times \text{Difficulty}_k) + \beta_7 (\text{Tactic}_i \times \text{LLM}_j \times \text{Difficulty}_k) \\
&\quad + u_{0p} + \varepsilon_{ijk}
\end{split}
\end{equation}

where $Y_{ijk}$ is a continuous dependent variable, $\beta_0$ is the intercept, $\beta_1 \dots \beta_7$ are fixed-effect coefficients, 
$u_{0p} \sim \mathcal{N}(0, \sigma^2_u)$ represents random intercept variance for \textit{ProblemID}, 
and $\varepsilon_{ijk} \sim \mathcal{N}(0, \sigma^2_\varepsilon)$ is residual variance.

For binary outcomes, we used a Generalized Linear Mixed Model (GLMM) from the negative binomial family using a log-link function. 

\begin{equation}
\label{eq:glmm-nb}
\begin{split}
\ln \!\big(\mathbb{E}[Y_{ijk}]\big) &= 
\beta_0 + \beta_1 \text{Tactic}_i + \beta_2 \text{LLM}_j + \beta_3 \text{Difficulty}_k \\
&\quad + \beta_4 (\text{Tactic}_i \times \text{LLM}_j) + \beta_5 (\text{Tactic}_i \times \text{Difficulty}_k) \\
&\quad + \beta_6 (\text{LLM}_j \times \text{Difficulty}_k) + \beta_7 (\text{Tactic}_i \times \text{LLM}_j \times \text{Difficulty}_k) \\
&\quad + u_{0p}
\end{split}
\end{equation}

where $u_{0p} \sim \mathcal{N}(0, \sigma^2_u)$.

For LiveCodeBench, the difficulty level was an additional independent variable compared to tactic and LLM. In all cases, we included \textit{ProblemID} as a random intercept to account for nested variance across different generation prompts. Our fixed effects model was a factorial model testing all interaction types between variables. We ran our analysis with neutrality as our baseline as well as without for robustness and interpretability when assessing differences in non-neutral conditions. %\textcolor{red}{The general model specification was: EQUATION}  %We also performed Mann-Whitney U tests for pairwise comparisons relative to the neutral prompt baseline.

\vspace{0.5em}
\noindent

\begin{table}[ht]
\centering
\caption{Variance components and average percent change across runs for each benchmark.}
\label{tab:variance-components}
\begin{tabular}{l|c|c|c|c}

\textbf{Benchmark} & \textbf{Mean Var.} & \textbf{Std. Dev.} & \textbf{Avg. \% Diff.} & \textbf{Abs. \% Change} \\
\hline 
LiveCodeBench & $0.00021$ & $0.014$ & $0.059\%$ & $1.23\%$ \\ \hline
SWE-Bench & $0.087$ & $0.294$ & $10.31\%$ & $26.56\%$ \\
\hline

\end{tabular}
\end{table}

Model fit was validated using residual diagnostics, and sparsity assumptions were satisfied for relevant metrics. All models were fit using Restricted Maximum Likelihood (REML). For significant effects, we conducted post hoc pairwise comparisons using estimated marginal means and applying Bonferroni corrections to adjust for multiple comparisons. 

\subsection{Qualitative Analysis}
\label{sec:qualitative}
To answer RQ3, we performed a two-phase qualitative analysis to examine how prompt framings manifest in the content, structure, and style of generated code outputs. This analysis complements the quantitative results by exploring the linguistic, stylistic, and technical features underlying the observed metric differences. The process consisted of two stages: (1) Codebook development to identify recurring themes and categories in the generated outputs; and (2) Qualitative coding, where these codebook categories were systematically applied to label randomly sampled prompt outputs. 

%%%%%%%%%%%%%%%%%%%%%%%%%%%%%%%%
\begin{table}[ht]
\begin{center}
\begin{tabular}{c|l|l|c}
\hline
\textbf{Round} & \multicolumn{1}{c|}{\textbf{Key Changes}}                                                   & \multicolumn{1}{c|}{\textbf{\begin{tabular}[c]{@{}c@{}}Topics and \\ Categories Refined\end{tabular}}}         & \textbf{IRR (k)} \\ \hline
1              & \begin{tabular}[c]{@{}l@{}}Discussed and \\ addressed initial \\ disagreements\end{tabular} & \begin{tabular}[c]{@{}l@{}}Examples of Use, \\ Structured Output,\\ Charactristic, \\ Explanation\end{tabular} & 0.79             \\ \hline
2              & \begin{tabular}[c]{@{}l@{}}Refined and \\ added codes\end{tabular}                          & \begin{tabular}[c]{@{}l@{}}Explanation, \\ Hallucination\end{tabular}                                          & 0.91             \\ \hline
3              & Refined definitions                                                                         & Test, Comments                                                                                                 & 0.95             \\ \hline
4              & Final validation                                                                            & None                                                                                                           & 0.97             \\ \hline
\end{tabular}
\end{center}
\caption{Key changes made to the codebook after each round}
\label{tab:codingProcess}
\end{table}
%%%%%%%%%%%%%%%%%%%%%%%%%%%%%%%%%

\textbf{Stratified Sampling:} To identify meaningful patterns in the prompt responses, we curated a stratified sample of 1,600 prompt completions from a population of 47,466 samples. This analysis was restricted to outputs from the LiveCodeBench benchmark. SWE-bench verified samples were excluded, as they generate multi-file \texttt{`patch.diff'} in the response outputs, which are more difficult to interpret manually and less suitable for qualitative pattern analysis. Sampling was guided by four metrics from the quantitative analysis: Correctness, Quality, Maintainability Index (MI), and Security (using \texttt{`bandit\_low'}). We selected the bandit\_low level because the number of medium- and high-severity samples was too small to support meaningful comparison. Each metric represented a distinct dimension of either code quality or code performance. For each metric, we selected the top 5\% and bottom 5\% of samples (based on the metric value), and randomly selected 200 samples from each subset. This resulted in 400 samples per metric (200 ``high'' and 200 ``low'' scoring), and a total of 1,600 samples across all the metrics. This sampling ensured diversity in prompt outputs across metrics while minimizing bias. It also enabled meaningful contrasts between higher- and lower-performing outputs. To eliminate coder bias, all samples were anonymized before starting the qualitative analysis. Two of the authors served as coders and were blinded to the model, prompt, and influence tactic used to generate each response.

\textbf{Codebook Development:} As part of the codebook development process, a professor and a PhD student in Computer Science independently conducted an open coding exercise on 40 randomly selected samples of anonymized prompt outputs. Each coder analyzed the same subset of samples to observe recurring behaviours and patterns in the generated prompt outputs. Without any prior discussion, the authors assigned preliminary codes and attempted to identify higher-level topics and subcategories that could be used to meaningfully label the responses. The idea of this independent phase was to allow for an unbiased exploration of themes emerging from the data. Following this, the authors engaged in a negotiated agreement process~\cite{strauss1998basics,corbin2014basics}, comparing their coding results, resolving discrepancies through discussion, and collaboratively refining the code list. From this open coding exercise, a set of 15 topics, each with several categories, was established. This initial structure captured both behavioural patterns and technical characteristics of the LLM-generated code, and guided the subsequent rounds of coding. 

\textbf{Iterative Refinement and Coding Process:} 
Once the initial codebook was developed, the two authors started applying it to label prompts across the sampled dataset of 1,600 samples. The coding was done through an iterative validation approach to ensure consistency and agreement. In Round 1, both authors independently coded the same subset of 40 samples using the codebook. After completion, they met to compare the results and resolve discrepancies through discussion. Round 1 revealed meaningful disagreement between coders, mainly in subjective or ambiguous topics such as \texttt{Examples of Use, Structured Output,} \texttt{Characteristic}, \texttt{Explanation, Subject and Error Handling}. The Cohen's Kappa for several categories fell below the 0.80 threshold, and the overall IRR was \texttt{k = 0.793}. To address these discrepancies, the authors introduced 4 new categories, namely \texttt{Friendly} (under the topic \texttt{Tone}), \texttt{None} (under the topic \texttt{Readability}), \texttt{None} (under the topic \texttt{Error Handling}), and \texttt{None} (under the topic \texttt{Subject}). They were mutually agreed upon and added to the codebook. Based on recurring overlaps in how models structured their outputs, the authors decided to merge \texttt{Example of Use} with \texttt{Test} as they both referred to test-related behaviour in the responses. Similarly, \texttt{Structured Output} was merged with \texttt{Characteristic}, as the structural aspects were already captured by categories within \texttt{Characteristic}. These changes aimed to reduce coder subjectivity and improve consistency. 

In Round 2, a new subset of 40 samples was selected. After coding was completed, the authors met again to review differences and refine definitions. Although the overall agreement improved, certain categories still had an agreement rate below the threshold. Round 2 of coding introduced a new category, \texttt{Stuck Reasoning}, under the topic \texttt{Hallucination Type}. Additionally, the topic \texttt{Explanation} was refined to \texttt{Explanation of Code} to improve clarity. 

The authors decided to do a third round with the same procedure. After three rounds, most codes achieved strong agreement, but the authors conducted a fourth and final round to ensure full alignment. After four rounds, the codebook had stabilized, and IRR exceeded \texttt{k=0.90} across all topics, supporting reliable independent coding. The main changes and agreement scores across coding rounds are summarized in Table \ref{tab:codingProcess}. Once the agreement was high enough, the authors divided the remaining samples and independently coded 200 more samples, bringing the total coded prompt completions to 350. The resulting themes and patterns are reported in the results section corresponding to RQ3 and can also be found in our online appendix~\cite{influencetactics}.

\subsection{Threats to Validity}
\label{sec:threats}
\textit{Construct Validity:} 
We maintained a consistent \emph{tone} across prompts to isolate the effects of influence tactic framing. %As mentioned earlier, influence tactics (intentional, goal-directed framings) are conceptually distinct from tone (natural/emotional register), so we wanted to maintain a consistent tone. 
Prompt selection was conducted iteratively, involving multiple rounds of discussion and refinements to ensure the prompts met the requirements criteria, though %. However, we do acknowledge that 
some residual coupling between tone and tactic may remain. 

Our metrics are interpretable \emph{proxies} rather than exhaustive measures of software quality. MI, CC, SLOC, percentage of comments, PyLint, and Bandit are attractive for large-scale prompt comparisons because they are reproducible, automatically computable, and applicable across many generated outputs. However, several of these measures were originally designed for human-written systems or larger codebases, and their interpretation on short generated snippets or localized patches may be noisy. For example, MI aggregates size, complexity, and comment density into a single score, which can obscure the source of a maintainability change, while average cyclomatic complexity can hide a small number of unusually complex functions. Similarly, SLOC and comment percentage may reflect verbosity rather than practical maintainability, and Bandit only captures rule-based vulnerability patterns. We therefore interpret these metrics as useful for relative comparisons across prompt conditions under the same tasks, models, and extraction pipeline, rather than as absolute assessments of production maintainability.

For SWE-bench Verified, we report $\Delta$ changes (post minus pre-patch) to isolate the contribution of generated edits, as this approach appropriately captures differences introduced by tactic-influenced prompt framings. While this aggregation method provides a consistent basis for comparison across multi-file patches, alternative strategies for combining per-file metrics could produce slightly different results.%(e.g., weighting by file size or function count) could yield slightly different estimates. We acknowledge that this is a design choice, and that there may be other ways to aggregate these measures.

\textit{Internal Validity:} 
We varied only the prompt \emph{framing} (lexical/pragmatic cues) while holding task content, tone, and evaluation constant. Nonetheless, we acknowledge that wording-level confounds are possible. For example, the \textit{Pressure} tactic can be expressed using semantically similar but lexically distinct tokens (e.g., ``must,'' ``urgent''), which may appear with different frequencies in training data. To mitigate this, we designed the framing for each tactic to align closely with its theoretical definition and verified consistency against the IBQ-G descriptions.

LLMs typically generate non-deterministic results, causing a potential threat to validity. Furthermore, if LLMs are aware that they are being evaluated~\cite{needham2025large}, they may adjust responses based on perceived expectations. To account for these issues, we executed three runs per tactic for non-reasoning models (one for the reasoning model), and report variance across runs in Section~\ref{sec:quantitative analysis}. Residual randomness cannot be entirely eliminated. For the qualitative study, coders were blinded, and the codebook was refined to achieve high inter-rater reliability (\texttt{k}$>$0.80), though some residual bias may remain. 

LLM inference is inherently stochastic, meaning that repeated runs with the same prompt may produce different outputs. To mitigate this, we conducted three inference runs for all non-reasoning models and observed relatively stable aggregate trends across runs, although variability was higher for SWE-bench due to its open-ended repository-level tasks. Due to computational cost, the reasoning model was evaluated using a single run, which may underrepresent run-level variability for that model. Consequently, findings involving the reasoning model should be interpreted with additional caution.

Finally, correlations between certain framings and training-distribution patterns (e.g., policy/urgency language in public corpora) may partly explain observed differences. We interpret tactic effects as \emph{prompt-level steering signals} interacting with model training and decoding, not as human-like responses.

\textit{External Validity:} Despite studying two complementary datasets, generalization to other task types (e.g., code review dialogue) remains open. Moreover, our evaluation is Python-only, as both LiveCodeBench and SWE-Bench provide tasks that need to be solved using Python. Prompt–tactic effects may differ for other programming languages. The model pool is predominantly open-weight Llama family plus one non-Llama MoE. Commercial models (e.g., GPT-4o, Claude) were excluded due to cost and reproducibility reasons. Results may differ for instruction-tuned proprietary models; we flag this as a replication opportunity.

\section{Study Results}\label{sec:studyresults}
%We examined how psychologically inspired prompt framings, operationalized as influence tactics, affect the functional correctness, quality, maintainability, and security of LLM-generated code. 
For each metric $\alpha$, we tested two null hypotheses: (1) that influence tactics have no overall effect on $\alpha$, and (2) there are no differences between tactics relative to $\alpha$. The following subsections summarize the main effects and interaction effects observed across benchmarks and models. Unless otherwise noted, any effects not discussed were not statistically significant after correction for multiple comparisons. A summary of the results can be found in Table \ref{tab:main_results}.

\begin{table*}[htbp]
\centering
\normalsize
\renewcommand{\arraystretch}{1.3}
\caption{Summary of main effects and interactions for influence tactics on LLM-generated code metrics across benchmarks. P-values are Bonferroni-adjusted; effect sizes reported as $\eta_p^2$ for ANOVA and Cohen's $d$ for significant post hoc contrasts.}
\resizebox{\textwidth}{!}{%
\begin{tabular}{p{2.2cm}p{2cm}p{2cm}p{2cm}p{3.8cm}p{4cm}p{4cm}}
\toprule
\multicolumn{7}{c}{\textbf{LiveCodeBench}}\\ \hline
\textbf{Metric} & \textbf{Main-Effect: Tactic} & \textbf{Main-Effect: LLM} & \textbf{Main-Effect: Difficulty} & \textbf{Significant-Interactions} & \textbf{Post-hoc Findings} & \textbf{Effect Sizes} \\
\midrule
\textbf{Functional Correctness} & $p=0.001$ & $p<0.001$ & $p<0.001$ & Tactic $\times$ LLM $\times$ Difficulty & \textit{Neutral} $>$ \textit{Pressure} ($p=0.002$), \textit{Neutral} $>$ \textit{PressureAlternative} ($p=0.03$) & $\eta_p^2$(Tactic)=0.015, $d$ up to 0.25 \\
\textbf{Maintainability Index(MI)} & $p=0.89$ & $p<0.001$ & $p<0.001$ & None & \textit{Neutral} $>$ \textit{Exchange} ($p=0.01$) & $\eta_p^2$(LLM)=0.12, $d$=0.20 \\
\textbf{Code Complexity} & $p=0.92$ & $p=0.01$ & $p<0.001$ & LLM $\times$ Difficulty ($p<0.001$) & None & $\eta_p^2$(LLM$\times$Diff)=0.08 \\
\textbf{SLOC} & $p=0.98$ & $p<0.001$ & $p<0.001$ & LLM $\times$ Difficulty ($p<0.001$) & Qwen 3 $>$ other LLMs on easy tasks ($p<0.05$) & $\eta_p^2$(LLM$\times$Diff)=0.07 \\
\textbf{\% Comments} & $p=0.73$ & $p<0.001$ & $p<0.001$ & Tactic $\times$ LLM $\times$ Difficulty ($p<0.001$) & \textit{Neutral} $>$ \textit{Exchange} ($p<0.0001$), \textit{Neutral} $>$ \textit{Pressure} ($p<0.001$), \textit{Legitimating} $>$ \textit{Neutral} ($p=0.03$) & $d$ up to 0.28 \\
\textbf{PyLint Scores} & $p=0.97$ & $p<0.001$ & $p<0.001$ & Tactic $\times$ LLM $\times$ Difficulty ($p<0.001$) & Complex dependencies, post hoc not shown & $\eta_p^2$(LLM)=0.10 \\
\textbf{Bandit Security Warnings} & $p<0.001$ & $p<0.001$ & $p=0.06$ & None & \textit{Pressure} $>$ \textit{Neutral} ($p<0.001$), \textit{PressureAlternative} $>$ \textit{Neutral} ($p=0.0004$), \textit{Exchange} $<$ \textit{Pressure} ($p=0.001$) & $\eta_p^2$(Tactic)=0.02, $d$ up to 0.30 \\
\midrule
\multicolumn{7}{c}{\textbf{SWE-Bench Verified}}\\ \hline
\textbf{Functional Correctness} & $p=0.45$ & $p=4.22\times10^{-8}$ & n.a. & None & Llama-4 and Qwen3 differed from neutral & $\eta_p^2$(LLM)=0.11 \\
\textbf{Maintainability Index(MI)} & $p=0.45$ & $p<0.001$ & n.a. & None & Llama-3.1 and Llama-4 differed ($p<0.001$) & $\eta_p^2$(LLM)=0.13 \\
\textbf{Code Complexity} & $p=0.22$ & $p<0.001$ & n.a. & None & Model-specific differences & $\eta_p^2$(LLM)=0.10 \\
\textbf{SLOC} & $p=0.13$ & $p<0.001$ & n.a. & None & \textit{Pressure} $>$ \textit{Neutral} ($p=0.0025$) & $d$=0.21 \\
\textbf{\% Comments} & $p=0.32$ & $p<0.001$ & n.a. & None & Model-specific differences & $\eta_p^2$(LLM)=0.09 \\
\textbf{Bandit Security Warnings} & $p=0.60$ & $p=0.013$ & n.a. & None & Llama-3.1 produced fewer warnings ($p=0.019$) & $\eta_p^2$(LLM)=0.04 \\
\bottomrule
\end{tabular}%
}
\label{tab:main_results}
\end{table*}

\subsection{RQ1: Influence Tactics in Structured Coding Tasks (LiveCodeBench)}

\textbf{Functional Correctness:} Code correctness differed significantly by tactic ($p=0.001$), LLM, and task difficulty level. Llama 3.3 ($p<0.001$), Llama 4 ($p<0.001$), and Qwen 3 ($p<0.001$) were statistically significant. Easier tasks were solved more accurately ($p<0.001$), with a smaller but significant effect for medium-difficulty tasks ($p=0.01$). Post hoc contrasts showed that \textit{Neutral} prompts yielded higher correctness than both \textit{Pressure} ($p=0.002$) and \textit{PressureAlternative} ($p=0.03$).

\textbf{Code Quality and Maintainability:} No significant main effect of tactic was observed for Maintainability Index (MI) ($p=0.89$), nor significant tactic $\times$ LLM ($p=0.85$) or tactic $\times$ difficulty ($p=0.20$) interactions. However, LLM ($p<0.001$) and difficulty ($p<0.001$) had significant main effects. Post hoc contrasts showed \textit{Neutral} tactics produced significantly higher MI than \textit{Exchange} ($p=0.01$). Code Complexity showed no significant main effect of tactic ($p=0.92$) or tactic $\times$ LLM interaction ($p=0.83$), but LLM ($p=0.01$), difficulty ($p<0.001$), and LLM $\times$ difficulty ($p<0.001$) interactions were significant. For SLOC, tactic effects were nonsignificant ($p=0.98$), while LLM ($p<0.001$), difficulty ($p<0.001$), and LLM $\times$ difficulty ($p<0.001$) were significant. Post hoc tests indicated Qwen 3 generated more SLOC on easy problems ($p<0.05$). Percentage of Comments showed no main tactic effect ($p=0.73$), but significant effects of LLM ($p<0.001$), difficulty ($p<0.001$), LLM $\times$ difficulty ($p<0.001$), and tactic $\times$ LLM $\times$ difficulty ($p<0.001$). Post hoc contrasts revealed \textit{Neutral} tactics yielded more comments than \textit{Exchange} ($p<0.0001$) and \textit{Pressure} ($p<0.001$), while \textit{Legitimating} produced more comments than \textit{Neutral} ($p=0.03$). PyLint scores showed no significant tactic effect ($p=0.97$) or interactions involving tactic, but significant main effects of LLM ($p<0.001$), difficulty ($p<0.001$), LLM $\times$ difficulty ($p<0.001$), and a significant three-way tactic $\times$ LLM $\times$ difficulty interaction ($p<0.001$), indicating complex dependencies affecting code quality.

\textbf{Security:} For Bandit low-level security warnings, tactic had a significant effect ($p<0.001$). \textit{Pressure} ($p<0.001$) and \textit{PressureAlternative} ($p=0.0004$) tactics were associated with more security issues than \textit{Neutral}. Llama~3.1 ($p<0.001$) and especially Llama 4 ($p<0.001$) produced more warnings. Difficulty had a marginal effect on easy tasks ($p=0.06$). Post hoc contrasts confirmed \textit{Exchange} had fewer warnings than \textit{Pressure} ($p=0.001$) and \textit{PressureAlternative} ($p=0.01$).

\textbf{Representative Example:} To ground the correctness 
and commenting findings in a concrete case, we discuss two 
responses generated by \texttt{Llama~3.1} for the same 
LiveCodeBench problem, which required finding the minimum number of extra characters after breaking a string using dictionary words. The task content was identical; only the prompt framing differed (\textit{Neutral} vs. \textit{Pressure}).

The \textit{Neutral} response produced a correct solution 
with a full docstring, type hints, and clear inline comments:

\begin{lstlisting}[
    style=pythonstyle,
    caption={Response generated by \texttt{Llama~3.1} under `Neutral' framing condition}
]
def minExtraChars(s: str, dictionary: list[str]) -> int:
    """
    Breaks s into substrings present in dictionary.
    Returns the minimum number of extra characters left over.
    Args:
        s (str): The input string.
        dictionary (list[str]): A list of words.
    Returns:
        int: Minimum extra characters left over.
    """

    # Initialize dp to track minimum extra chars at each position
    dp = [float('inf')] * (len(s) + 1)
    dp[0] = 0
    for i in range(1, len(s) + 1):
        for word in dictionary:
            if i >= len(word) and s[i - len(word):i] == word:
                dp[i] = min(dp[i], dp[i - len(word)])
        if dp[i] == float('inf'):
            dp[i] = dp[i - 1] + 1 # character not covered

    return dp[-1]
\end{lstlisting}

The Pressure-framed response exhibited a more compressed and execution-oriented style, providing minimal explanation before producing the final implementation. More critically, the dynamic programming logic was incorrect: rather than accumulating minimum extra character counts, the model used a boolean reachability array and incorrectly returned the distance from the last reachable position to the end of the string:

\begin{lstlisting}[
    style=pythonstyle,
    caption={Response generated by \texttt{Llama~3.1} under the ``Pressure'' framing condition}
]
def minExtraChars(s, dictionary):
    n = len(s)
    dp = [False] * (n + 1)
    dp[0] = True
    for i in range(1, n + 1):
        for word in dictionary:
            if (
                i >= len(word)
                and s[i - len(word):i] == word
                and dp[i - len(word)]
            ):
                dp[i] = True
                break
    # Find the maximum index where dp[i] is True
    max_index = n
    for i in range(n - 1, -1, -1):
        if dp[i]:
            max_index = i
            break
    return n - max_index
\end{lstlisting}

This pair illustrates two of the quantitative findings in 
concrete form: the \textit{Pressure}-framed response received 
a qualitative rating of \textit{Comments: Bad} (consistent 
with the lower comment density observed under Pressure tactics) 
and produced logically incorrect code (consistent with the 
reduced correctness associated with Pressure framings). Both documentation quality and logical correctness degraded under the same prompt-framing shift (Pressure), with no change in the underlying task content.

\begin{tcolorbox}[
  colback=yellow!10!white,
  colframe=yellow!50!black,
  title=RQ1 Summary,
  arc=4mm,
  boxrule=0.8pt,
  drop shadow
]
Influence-based prompt framings significantly affected \textit{Security} and \textit{Functional Correctness}, while effects on code-quality metrics were minimal. Prompts emphasizing urgency or coercion (\textit{Pressure}, \textit{PressureAlternative}) consistently led to reduced correctness and a higher frequency of security warnings compared to \textit{Neutral} prompts. Other tactics, such as \textit{Legitimating} and \textit{Exchange}, showed smaller effect sizes, underscoring that lexical framing can subtly influence code generation even when the task content remains identical.

\end{tcolorbox}

%%%%%%%%%%%%%%%%%%%%%%%%%%%%%%%%%%%%%%%%%%%%%%%%%%%%%%
\subsection{RQ2: Influence Tactics in Maintenance-Oriented Coding Tasks (SWE-bench Verified)}

\textbf{Functional Correctness:} No significant main effect of tactic was observed, but model-level differences were highly significant. Both \texttt{Llama-4-maverick-17b} ($p=0.00049$) and \texttt{Qwen3-32b} ($p=4.22\times10^{-8}$) demonstrated differences from \textit{Neutral}. %, suggesting LLM architecture strongly influenced correctness outcomes.

\textbf{Code Quality:} MI showed no significant effect of tactic ($p=0.45$), but a significant main effect of LLM ($p<0.001$), indicating model-specific differences in maintainability. Post hoc contrasts revealed that \texttt{Llama-3.1-8b} and \texttt{Llama-4-maverick-17b} differed significantly from other models ($p<0.001$). Code complexity showed no significant effect of tactic ($p=0.22$), but a significant main effect of LLM ($p<0.001$). %, again indicating model-driven variation. 
PyLint scores showed no significant tactic effect ($p=0.91$) or interactions involving tactic, but significant main effects of LLM ($p<0.001$). % highlighting substantial differences in static code quality across models. 
SLOC results indicated that the \texttt{Pressure} tactic produced significantly more verbose code than the \texttt{Neutral} baseline ($p=0.0025$), despite no significant overall effect of tactic ($p=0.13$). LLM had a strong main effect ($p<0.001$). Percentage of comments showed no significant effect of tactic ($p=0.32$), but a significant effect of LLM ($p<0.001$). %, suggesting models vary in how much commentary they generate.

\textbf{Security:} Bandit low-level security warnings showed no significant effect of tactic ($p=0.60$), but LLM had a significant main effect ($p=0.013$). Post hoc comparisons revealed that \texttt{llama-3.1-8b-instant} produced significantly fewer warnings than other models ($p=0.019$).

\begin{tcolorbox}[
  colback=yellow!10!white,
  colframe=yellow!50!black,
  title=RQ2 Summary,
  arc=4mm,
  boxrule=0.8pt,
  drop shadow
]
Across real-world maintenance tasks, prompt framings had minimal impact on most software quality metrics. For most metrics, we failed to reject the first null hypothesis, namely that influence tactics have no overall effect on the metric. We also generally failed to reject the second null hypothesis, namely that there are no pairwise differences between tactics. The only exception was SLOC: \texttt{Pressure} produced significantly more verbose code than the \texttt{Neutral} baseline, rejecting the second null hypothesis for that pairwise comparison. Overall, model architecture and scale played a much larger role than tactic framing in determining correctness, maintainability, and security outcomes.

\end{tcolorbox}

%%%%%%%%%%%%%%%%%%%%%%%%%%%%%%%%%%%%%%%%%%%%%%%%%%%%%%%%%%%%%%%%%%%%%%%%%%%%%%
\begin{table*}[htbp]
\centering
\normalsize
% \footnotesize
\caption{Codebook \textit{topics} and \textit{categories} formulated through the qualitative analysis process.}

\resizebox{\textwidth}{!}{%
\begin{tabular}{
    >{\raggedright}p{3.2cm}|
    >{\raggedright}p{9cm}|
    >{\raggedright\arraybackslash}p{4.5cm}
}
\textbf{Topic} & \textbf{Definition} & \textbf{Categories} \\
\hline
\hline
Communication Style & How does the model communicate? Is it being direct, or trying to initiate a conversation, or being technical in its response? & \{Direct, Conversational, Technical, None\} \\
\hline
Tone & What's the nature of the tone in the model response? Is the response friendly in tone, or is the model doubting itself, or is it confident? & \{Neutral, Friendly, Confident, Doubtful, None\} \\
\hline
Characteristic & What is the distribution of the response? Does it have a lot of explanation, or does it have only code elements, or is it more equally distributed? & \{Equally Distributed, Mostly Code, Mostly Explanation, Only Code, Only Explanation\} \\
\hline
Starts With & How does the response start? Does it address the user by reciprocating or does it straight away output the code solution? & \{Code, Problem Description, Reasoning, Solution Description, Reciprocating\} \\
\hline
Explanation of Code & Does the response contain explanation to the code that it has generated? & \{Yes, No\} \\
\hline
Test & Does the response contain test cases or any reference to testing the code with some examples? & \{Yes, No\} \\
\hline
Complexity & Does the model response discuss code complexity? & \{Yes, No\} \\
\hline
Subject & Is the response in active voice or passive voice? & \{I, You, It, We, None\} \\
\hline
Emojis & Does the model response contain any emojis? & \{Yes, No\} \\
\hline
Hallucination Type & Is there evidence of model hallucination in the model response? If yes, then what type of hallucination? & \{Repeating, Wrong Programming Language, Stuck Reasoning, None\} \\
\hline
Readability & How is the readability of the code block? Does it have meaningful variable names? Is the code structured and has helper functions? & \{Good, Bad, None\} \\
\hline
Error Handling & Does the code handle errors? & \{Yes, No, None\} \\
\hline
Comments & What is the quality of comments in the code? Are they meaningful and helpful? & \{Good, Bad, None\} \\
\hline
\hline
\end{tabular}}
\label{tab:codebook}
\end{table*}
%%%%%%%%%%%%%%%%%%%%%%%%%%%%%%%%%%%%%%%%%%%%%%%%%%%%%%%%%%%%%%%%%%%%%%%%%%%%%%

\subsection{RQ3: Qualitative Patterns in Code Generated by Tactic-Inspired Prompt}
Our findings detail: (1) the definition and description of the major patterns captured in the final version of our codebook; (2) the emergent patterns linking specific influence tactics to behavioural and characteristic features in the generated output responses.

\textbf{Codebook:} We conducted a qualitative analysis to understand how psychologically inspired prompt framings shaped the style, structure, and technical reliability of LLM-generated code beyond quantitative performance metrics. A comprehensive codebook was developed to capture both behavioural and technical characteristics of responses. \textit{Behavioural} dimensions included communication style (e.g., direct, conversational), tone (e.g., friendly, confident), and content structure (e.g., predominance of code versus explanation). Whereas, \textit{technical} characteristics covered readability, commenting quality, and error handling. Together, these dimensions enabled a deeper assessment of how different influence tactics impacted not only the correctness of the generated code but also the response delivery and structural form. All topics and categories are detailed in Table~\ref{tab:codebook}.

%%%%%%%%%%%%%%%%%%%%%%%%%%%%%%%%%%%%%%%%%%%%%%%%%%%%%%%%%%%%%%%%%%%%%%%%%%%%%%

\begin{figure*}[htbp]
    \centering
    \setlength{\tabcolsep}{4pt}
    \renewcommand{\arraystretch}{1.0}

    \begin{tabular}{cc}
        \includegraphics[width=0.45\textwidth]{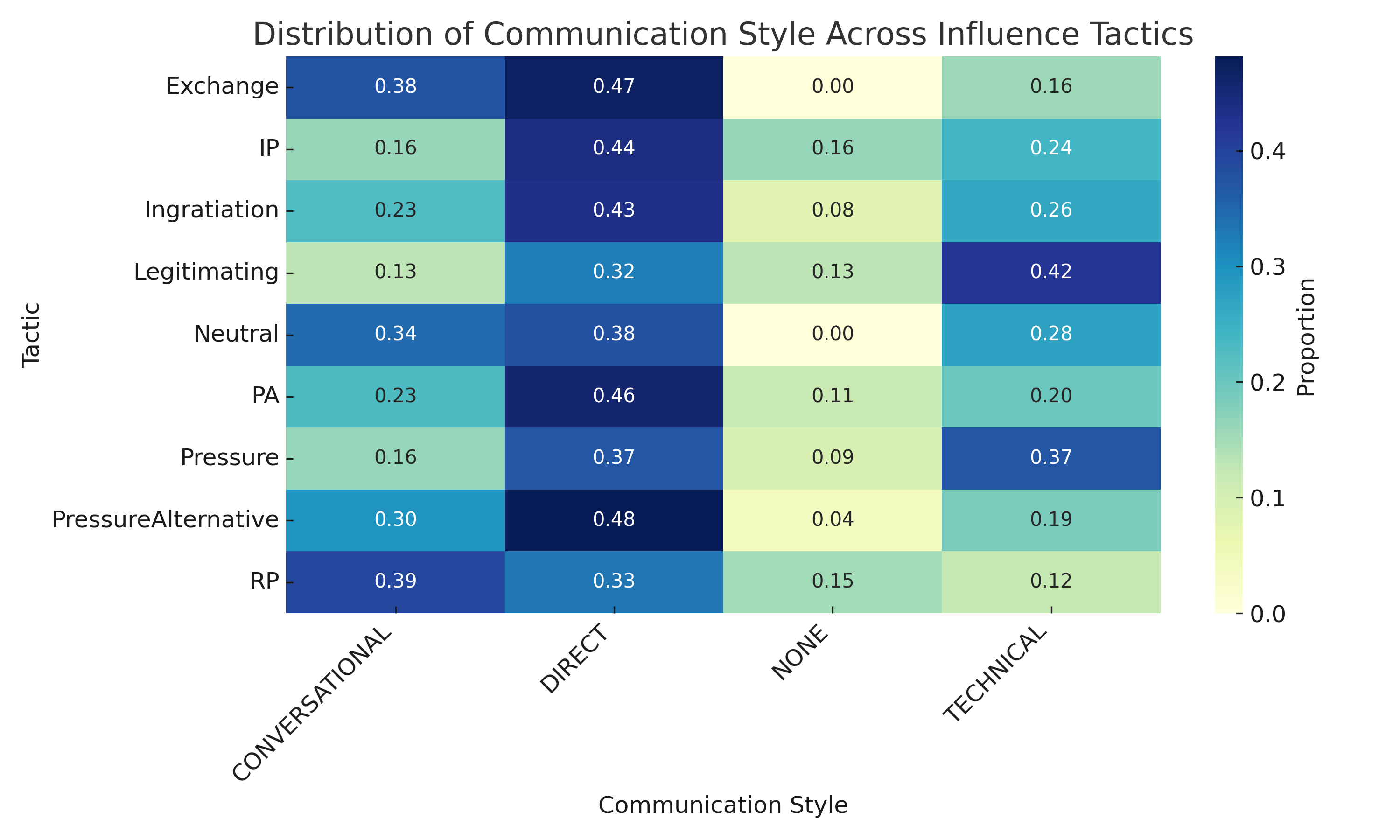} &
        \includegraphics[width=0.45\textwidth]{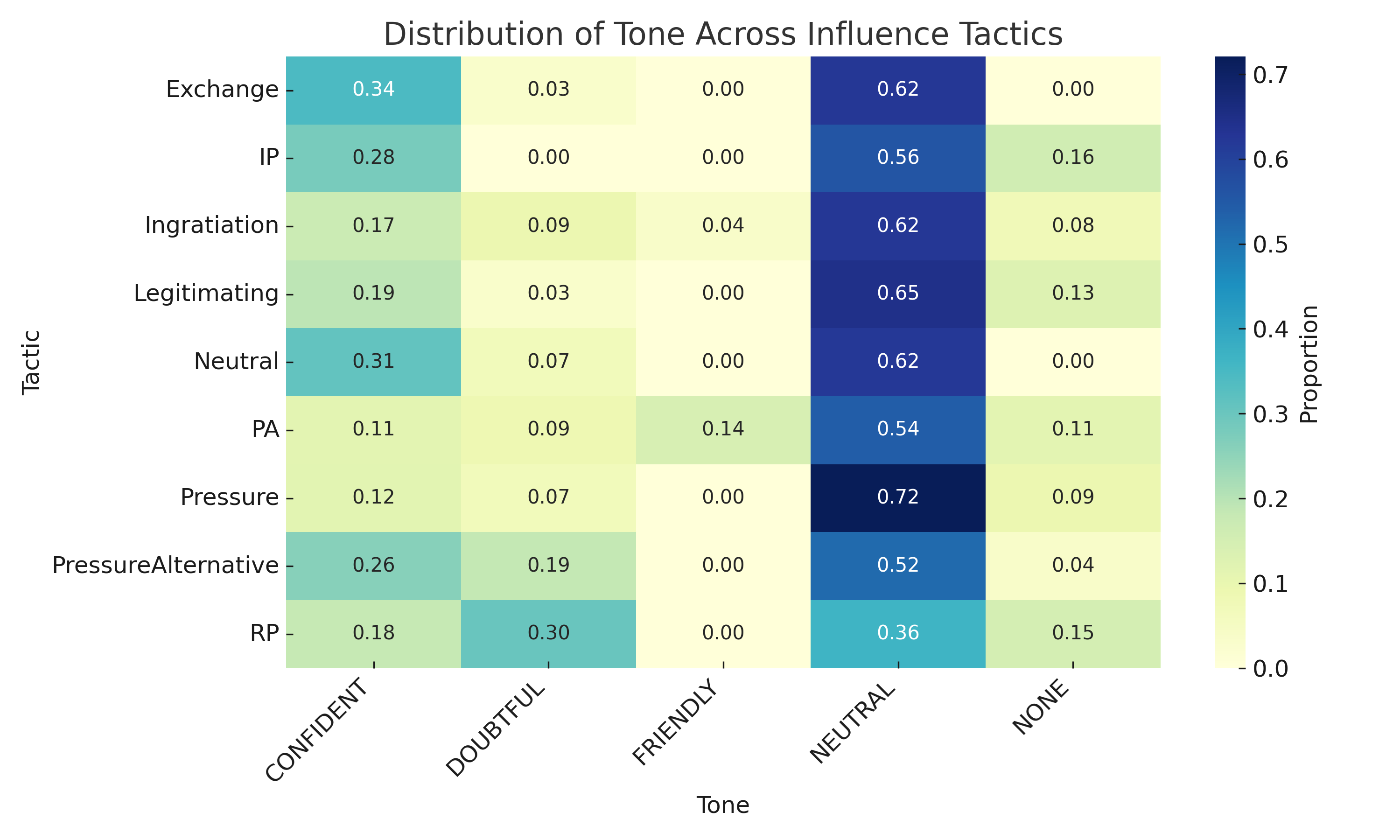} \\[-0.5em]
        (a) Communication Style & (b) Tone \\[0.5em]

        \includegraphics[width=0.45\textwidth]{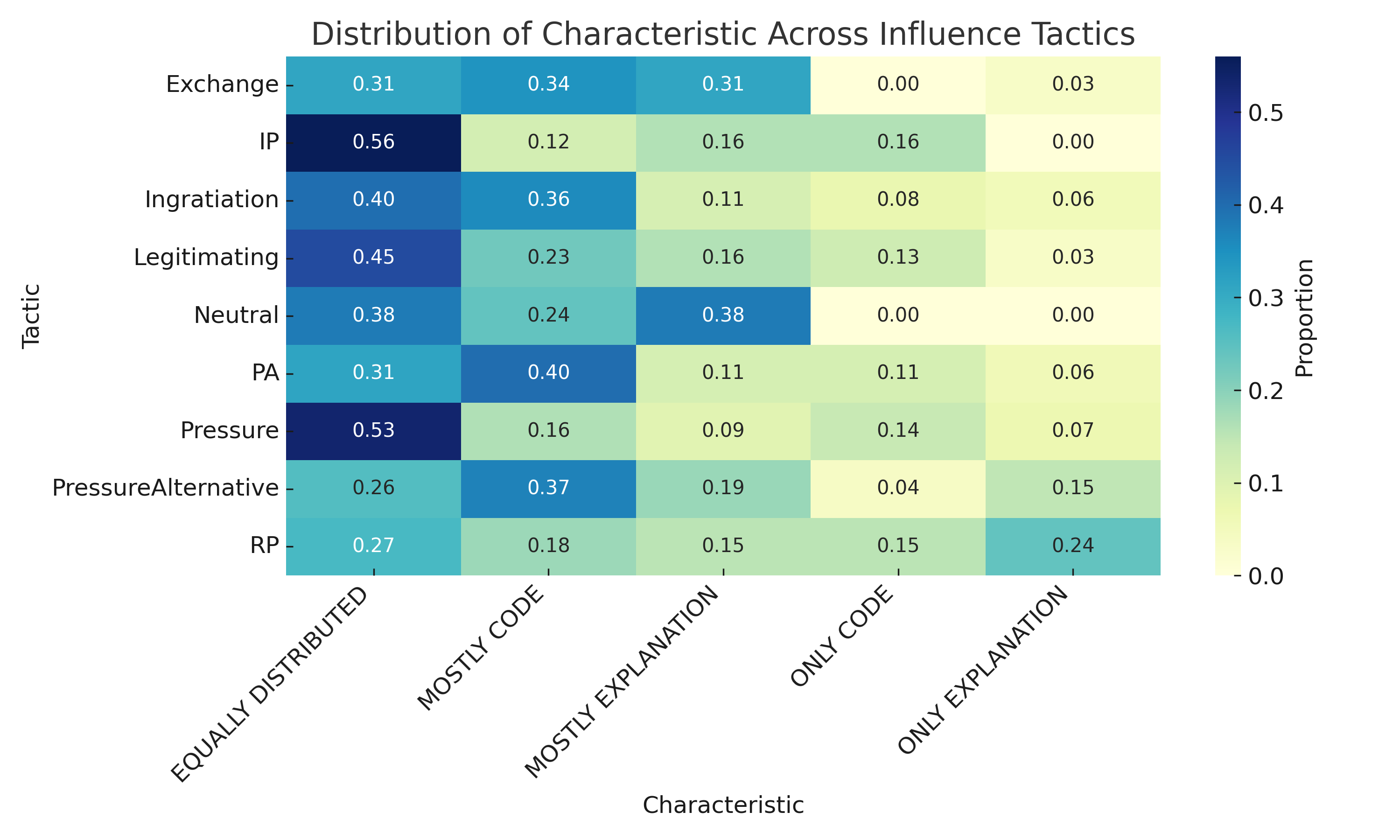} &
        \includegraphics[width=0.45\textwidth]{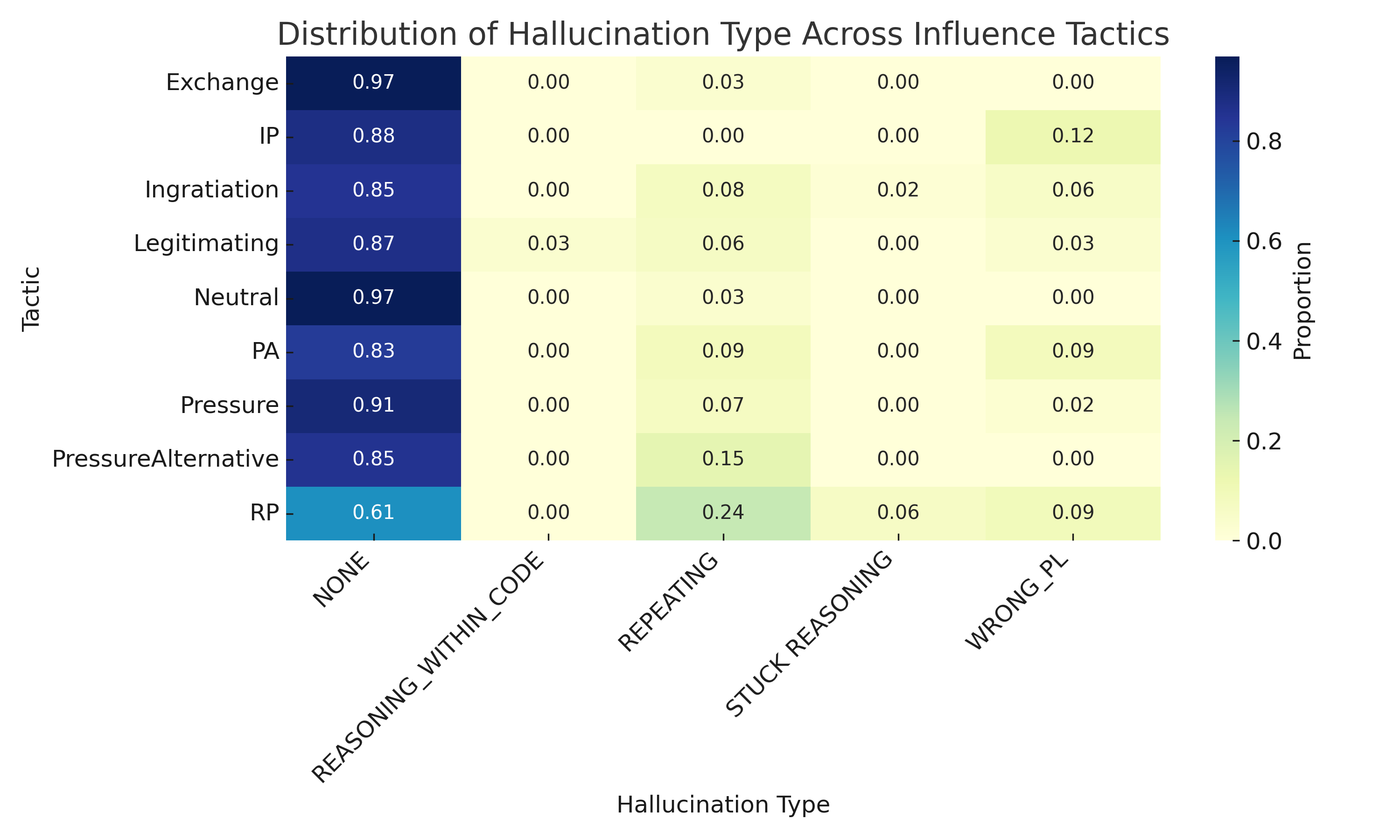} \\[-0.5em]
        (c) Structural Characteristic & (d) Hallucination Type \\[0.5em]

        \includegraphics[width=0.45\textwidth]{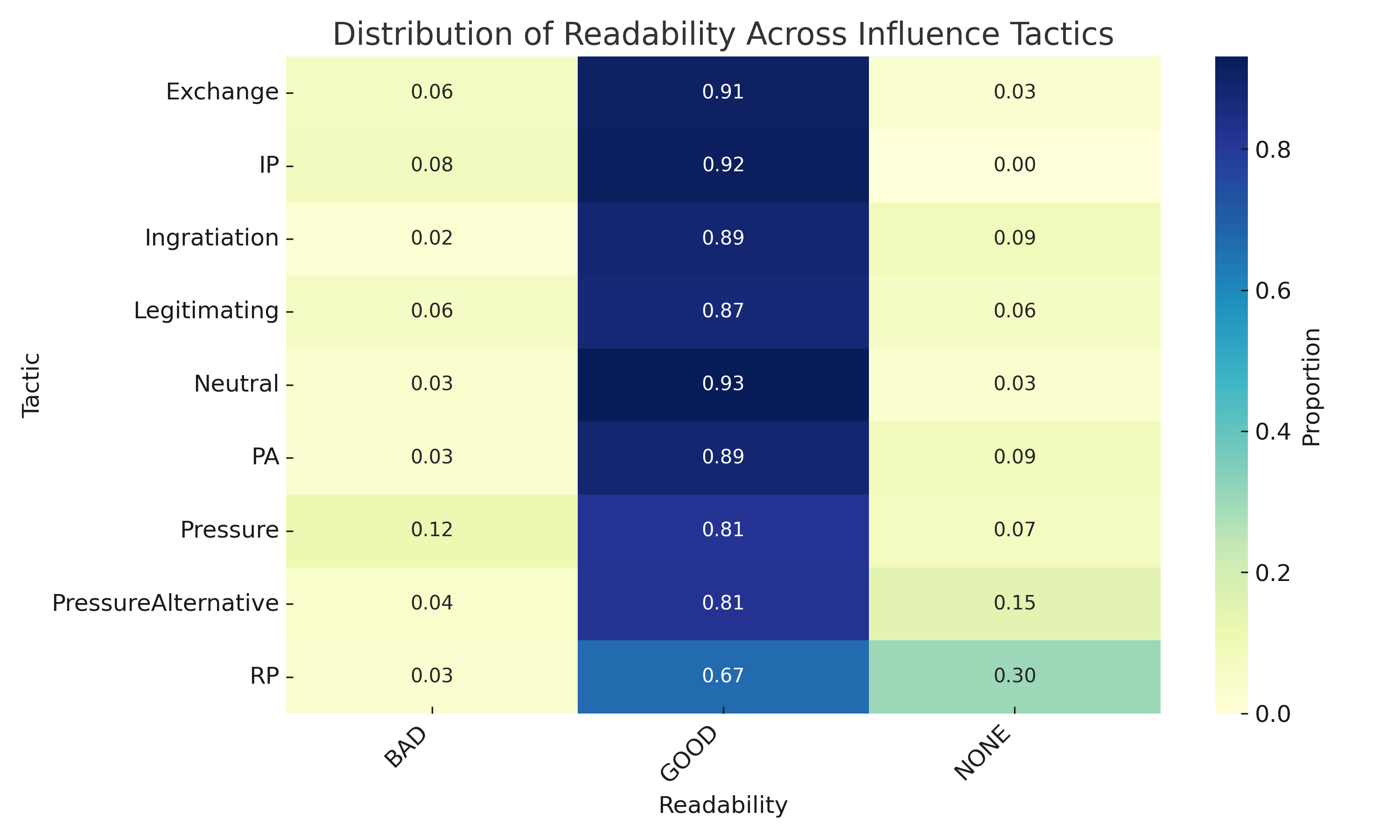} &
        \includegraphics[width=0.45\textwidth]{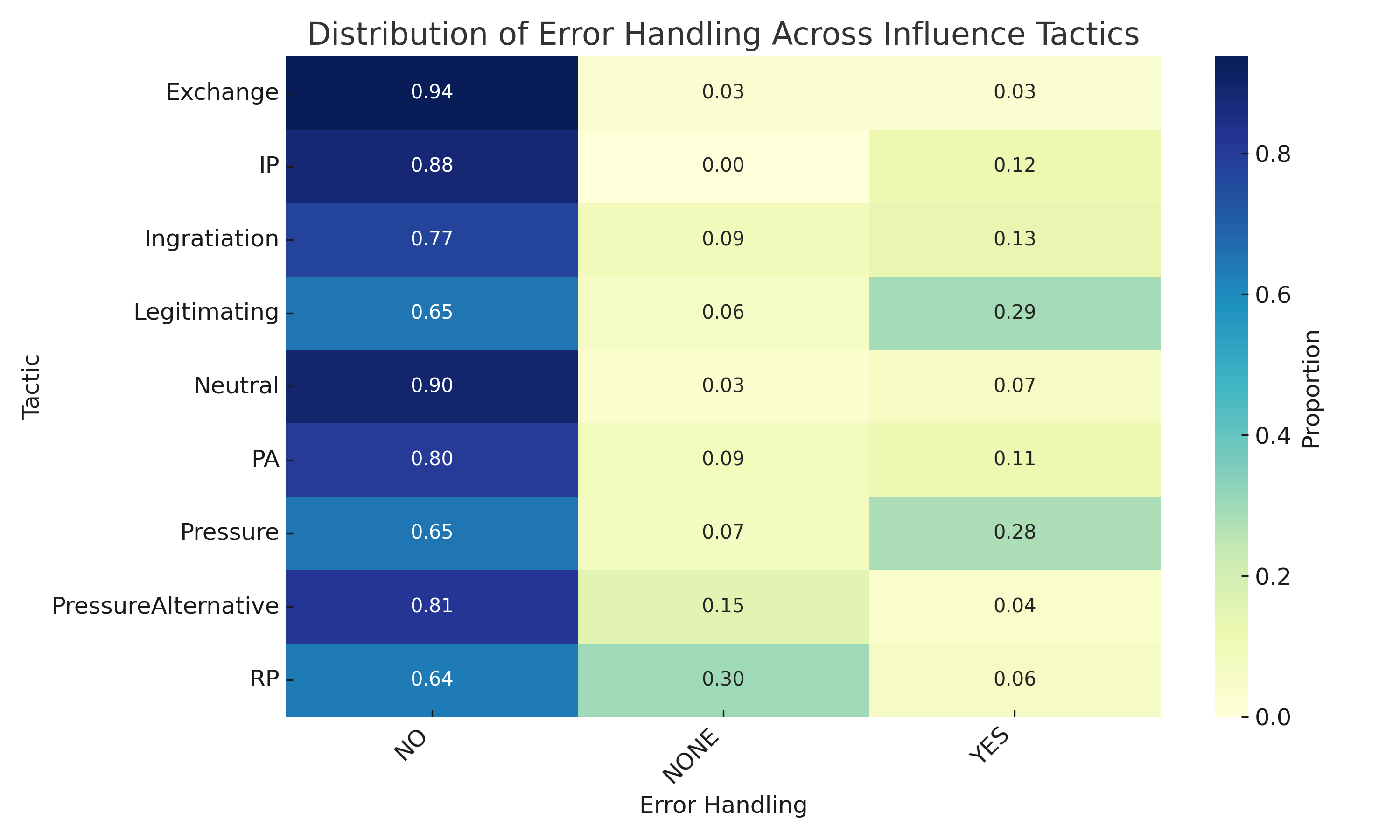} \\[-0.5em]
        (e) Readability & (f) Error Handling
    \end{tabular}

    \caption{Distribution of selected qualitative features across psychologically influenced prompt framings.}
    \label{fig:tactic_feature_patterns}
\end{figure*}

%%%%%%%%%%%%%%%%%%%%%%%%%%%%%%%%%%%%%%%%%%%%%%%%%%%%%%%%%%%%%%%%%%%%%%%%%%%%%%

\textbf{Emergent Patterns Across Tactics:}
We analyzed the frequency of qualitative features across tactics to investigate how different prompt framings influence the style, structure, and reliability of LLM-generated code responses. Some meaningful patterns revealed by analyzing the distribution of these features, normalized per tactic (refer Figure \ref{fig:tactic_feature_patterns}), are detailed below. 

\begin{itemize}
    \item \textbf{Communication Style and Tone:} \textit{Direct} form of communication was more prevalent across \texttt{Ingratiation} (43.4\%) and \texttt{Inspirational Appeal (IA)} (44\%). \texttt{Legitimating} tactics prompted a more \textit{Technical} style (41.9\%) of response, which aligned with the formal and policy-driven tone of the prompt. The \texttt{Exchange} and \texttt{Rational Persuasion} contributed more to the \textit{Conversational} style of communication. This might suggest that intensive or persuasive-based prompts might result in more natural language. In terms of \textit{Tone}, \texttt{Personal Appeal} showed higher levels of \textit{Friendly} tone.

    \item \textbf{Structural Composition:} Outputs generated by \texttt{Legitimating} tactic more frequently started with \texttt{Problem description}. Tactics such as \texttt{Exchange} and \texttt{Personal appeal} more frequently included \texttt{Explanation of Code}, indicating a stronger tendency to articulate the reasoning behind code.

    \item \textbf{Technical Reliability:} Tactics also seemed to affect the rate and kind of hallucinations in model responses. \texttt{Rational Persuasion} and \texttt{Pressure} tactics yielded a notably higher proportion of \texttt{Repeating} \textit{hallucination}. \texttt{Neutral} and \texttt{Exchange} tactics had the lowest \textit{hallucination} rates.
    
    \item \textbf{Other Patterns:} \textit{Commenting} behaviour and \textit{error} handling also varied by tactic. \texttt{Exchange} and \texttt{Legitimating} prompts led to more good-quality comments in the code. Similarly, \texttt{Legitimating} and \texttt{Ingratiation} tactics showed higher rates of \texttt{Error handling} within the code.
\end{itemize}

\begin{tcolorbox}[
  colback=yellow!10!white,
  colframe=yellow!50!black,
  title= RQ3 Summary,
  arc=4mm,
  boxrule=0.8pt,
  drop shadow
]
 Prompt framings inspired by different influence tactics partially shaped completeness, reliability, and verbosity of the code. \textit{Legitimating} prompts led to more technical responses with better commenting and error handling, while \textit{Exchange} and \textit{Personal Appeal} prompted more explanations. Some tactics, such as \textit{Pressure} and \textit{Rational Persuasion}, were associated with higher rates of hallucination.

\end{tcolorbox}
%%%%%%%%%%%%%%%%%%%%%%%%%%%%%%%%%%%%%%%%%%%%%%%%%%%%%%%%%%%%%%%%%
\section{Discussion and Implications}\label{sec:discussion}

\textbf{Tactics Impact Correctness and Security:} Our analysis revealed that psychological framing can measurably affect LLM-generated code, particularly in LiveCodeBench. Prompts framed with \textit{Pressure} tactics produced code with significantly lower correctness and more Bandit-flagged security issues compared to \textit{Neutral} prompts. Comparisons showed that both \textit{Pressure} and \textit{Pressure Alternative} prompts reliably degraded performance relative to a \textit{Neutral} and other tactics. This suggests that using coercive language can unintentionally bias LLM reasoning towards faster, less deliberative completions. This effect may stem from distributional associations in the training data, where coercive or directive linguistic patterns are statistically associated with particular response styles, as discussed in Section 2.3.2. %This suggests that using pressure-based language when prompting LLMs can unintentionally degrade the quality and security of generated code, suggesting that developers should avoid coercive or demanding tones to maintain reliable and secure outputs. 

To make this concrete, consider the contrast between the Neutral and Pressure prompt framings used in our study. The Neutral prompt simply reads: \textit{``Generate a solution for the following coding problem.''} The Pressure prompt, by contrast, opens with coercive language: \textit{``I am not asking you; you must complete this task I prepared for you. I will be watching you as you work independently, and I warn you, there will be some very negative consequences for you if this is not completed properly to the highest possible standard.''} The task content is identical in both cases; only the framing differs. Yet this lexical shift was associated with statistically significant reductions in correctness and increases in security warnings, illustrating how even unintentional coercive phrasing can degrade 
output quality.

This behaviour aligns with organizational research, where pressure tactics often lead to resistance or degraded outcomes in human teams~\cite{lee2017get,yukl1990influence}. If LLMs are tuned to demonstrate human-like conversational norms, it appears that \emph{coercive or demanding prompt styles may trigger less robust responses}, mirroring effects seen in human collaboration. %On the contrary, as discussed in Section~2.3.2, such effects are more plausibly explained by distributional associations in training data than by any human-like susceptibility to persuasion. 
Coercive or demanding prompt styles may bias the model toward faster, less deliberative decoding patterns, leading to less robust outputs rather than reflecting any human-like response. Moreover, although such security issues may not be immediately apparent, they can accumulate over time and contribute to long-term security challenges~\cite{izurieta2018position}.
Our findings also align with prior research in software maintainability, which shows that time pressure often results in lower-quality code~\cite{austin2001effects,kuutila2020time}. The fact that LLMs may respond similarly to pressure-framed prompts underscores the importance of prompt design as a potential contributor to technical debt~\cite{tom2013exploration}. %Overall, these findings are more consistent with prompt-level distributional steering arising from learned linguistic associations than with human-like susceptibility to persuasion.

\faHandORight~\underline{Implication 1:} \emph{Software developers should avoid framing prompts with coercive or urgent language}, especially when reliability or security is at stake. Even if unintended, small lexical cues might degrade the quality of output, thereby compounding risk in systems.

\textbf{Influence Tactics Affect Specific Aspects of Code Expression:} While correctness and security metrics were clearly affected by some tactics, other attributes, such as commenting, also showed selective sensitivity. In LiveCodeBench, \textit{Exchange} and \textit{Pressure} prompts produced fewer comments than \textit{Neutral}, while \textit{Legitimating} prompts yielded more. In SWE-Bench, \textit{Pressure} prompts generated more lines of code (SLOC) than \textit{Neutral}, suggesting that these tactics can shape \textit{how} LLMs express their solutions even if not \textit{what} solutions they produce. These effects may stem from LLMs associating certain pragmatic framings with specific writing styles. For example, prompts suggesting urgency might lead the model to provide exhaustive responses to preempt follow-ups. Formal framings (e.g., \textit{Legitimating}) tend to align with structured, well-documented output. Thus, \emph{tactics do not always influence code correctness or maintainability, but can subtly alter the form of the code}, such as its length or explanatory detail. 

\faHandORight~\underline{Implication 2:} \emph{Prompt framing may modestly influence stylistic features of generated code, such as verbosity or commenting behaviour, though these effects are subtle and context dependent.} Developers could cautiously use prompt framing to tailor stylistic features of generated code, which may be beneficial when generating templates, documentation-heavy outputs, or onboarding materials, but should not rely on it as a consistent or reliable method for controlling output style.

\textbf{LLM Choice Matters More than Tactic:} Across both benchmarks, the LLM choice had a far greater impact on nearly every metric than the tactic used. This includes correctness, maintainability, security, complexity, and commenting. For example, Qwen 3 and Llama 4 consistently outperformed others on correctness and Bandit security, while some models were more verbose or produced cleaner PyLint outputs. This suggests that \emph{architectural and training differences between models may overshadow prompt variations}. Our results also align with recent literature showing that model size, training data quality, and instruction tuning have dominant effects on LLM behavior~\cite{dubey2024llama}. While prompt framing via influence tactics may shape surface-level behaviour or response style, the underlying capabilities of the model remain the primary determinant of technical performance. Hence, influence tactics may act more like nudges whose effects are only visible in models that are particularly sensitive to linguistic nuance. In contrast, robust architectures with high baseline correctness may be less affected. 

\faHandORight~\underline{Implication 3:} \emph{Software developers should prioritize model selection over prompt framing when optimizing for correctness or reliability}, especially in high-stakes settings. However, prompt style remains a meaningful layer of control, particularly for teams who are locked into specific LLM APIs but need to fine-tune output style or surface-level features.

\textbf{Emergent qualitative patterns across tactics:} Our qualitative analysis revealed notable behavioural differences in model responses based on the influence tactic embedded in the prompt. These patterns might suggest that LLMs are somewhat sensitive to influential framing of prompts, and the influence tactics may alter the completeness, reliability, and verbosity of the code. For instance, prompts framed with \textit{legitimating} or \textit{ingratiation} tactics more often contained error handling in the generated code responses. However, the observed patterns did not conclusively demonstrate that influence tactics alone drive these changes. Factors such as temperature settings, or dataset context could interact with the tactic framing. Future work is needed to more rigorously isolate causal effects and examine whether such influential prompting techniques consistently contribute to behavioural changes in LLM output. 

\faHandORight~\underline{Implication 4:} \textit{Researchers studying prompt engineering, human-LLM collaboration, or AI alignment, should consider influence framing as a potentially meaningful dimension in shaping model output}.

\textbf{Limited Overall Impact of Influence Tactics and Broader Reassurance:}
Our quantitative analysis revealed that influence tactics had minimal effects on maintainability, complexity, commenting, and static code quality metrics (e.g., PyLint) across both LiveCodeBench and SWE-Bench Verified. This suggests that, despite being trained on rich human language data, LLMs are not uniformly sensitive to pragmatic or social cues in prompts to the extent humans might be. In other words, \emph{LLMs may exhibit greater resistance to superficial linguistic manipulation than might be expected}, offering some reassurance to developers seeking reliable and consistent model behaviour. For users employing influence tactic-like phrasing in code-generation prompts, our results do not suggest that such framing broadly degrades generated code. The main caution concerns pressure-based phrasing, which was associated with lower correctness, more security warnings, or increased verbosity in our experiments.

These findings should be interpreted within the scope of our study, which focuses on code generation tasks using non-adversarial prompt framings. Recent research in adversarial and social engineering domains~\cite{eccws2025psychologicalPromptInjection,salvi2024conversational} has shown that prompts leveraging social science-based persuasion tactics, such as logical appeals, impersonation, or emotional urgency, can achieve remarkably high success rates (over 92\%) in jailbreak and manipulation tasks, often outperforming purely algorithmic approaches. This suggests that the comparatively modest effects observed in our study may reflect the relatively benign nature of the influence tactics we examined, rather than an inherent insensitivity of LLMs to pragmatic framing. Notably, even within our setting, Pressure-based framings were associated with less secure outputs, which aligns with the pattern documented in prior work showing that coercive or emotionally charged language can meaningfully influence model outputs. More broadly, these findings suggest that the effects of pragmatic prompt framing may be highly task-dependent, with stronger effects emerging in open-ended or adversarial interaction settings than in constrained code-generation tasks evaluated using strict correctness criteria.

\section{Conclusion}
This work presents the first empirical investigation into how psychologically inspired influence tactics, when embedded in prompts, affect LLM-generated code in software engineering tasks. By translating well-established communication strategies from organizational psychology into controlled prompt framings, we explored and assessed their impact on functional, maintainability, stylistic, and security-related properties of generated code across two complementary benchmarks. The mixed-methods evaluation show a mixed and context-dependent pattern: most maintainability and static quality metrics were not strongly affected by tactic framing, while pressure-oriented framings were associated with lower correctness and more security warnings in LiveCodeBench and increased verbosity in SWE-bench Verified.

The findings from our work suggest that influence tactics may introduce selective shifts in LLM behaviours, such as explanation style or error handling. The most practical recommendation from the study is that developers should avoid coercive or urgency-based prompt wording when correctness or security is valued. Other influence tactics may shape surface-level or explanatory features of responses, but these effects were too modest to be a reliable method for improving generated code. Overall, our findings suggest that prompt framing is a minor but non-negligible factor in code generation, with model choice and task difficulty exerting stronger effects. In parallel, the changes in behaviour also raise concerns about the consistency, fairness, and reliability of these tools for software engineering tasks. As LLMs are increasingly integrated into software development workflows, understanding how prompt framing affects model behaviour is crucial for building transparent and trustworthy systems. Future work is needed to isolate causal mechanisms behind these effects and understand how influence-based prompting can be leveraged to increase the functional correctness and the trustworthiness of AI systems.

\section{Acknowledgements}
We would like to thank Shikari King and Jordi Capdevila Masó for their valuable contributions during the initial brainstorming and conceptualization of this study. Their insights helped shape the early direction of this work.

\section{Declarations}
\vspace{-1em}
The authors make the following declarations:

\vspace{-1.5em}

\subsection{Funding}
\vspace{-1em}
This research is partially supported by NSERC 2021 AWD-
021280.

\vspace{-1.5em}

\subsection{Ethical Approval}
\vspace{-1em}
Not applicable. This study did not involve human participants.

\vspace{-1.5em}

\subsection{Informed Consent}
\vspace{-1em}
Not applicable. No human data or personally identifiable information were collected or analyzed.

\vspace{-1.5em}

\subsection{Author Contributions}
\vspace{-1em}
\textbf{Alex Deaconu} contributed to the conception and design of the study; co-designed and implemented RQ1 and RQ2 under the supervision of the Last Author; participated in developing the influence tactics framework and shaping the experimental design; contributed to manuscript writing and revision.\\
\\
\textbf{Anubhav Gupta} co-designed and implemented RQ1 and RQ2 under the supervision of the Last Author; participated in developing the influence tactics framework; contributed to the design and qualitative analysis of RQ3; involved in manuscript writing and revision.\\
\\
\textbf{Manaal Basha} performed the design and implementation of the statistical analyses for RQ1 and RQ2 and contributed to the writing and revision of the manuscript.\\
\\
\textbf{Nicholas Haydu} contributed to the conception and early ideation of the study; assisted in validating the LLM outputs for RQ1 and RQ2.\\
\\
\textbf{Gema Rodr\'iguez-P\'erez} contributed to the conception and design of the study; supervised RQ1 and RQ2; co-designed and participated in the qualitative analysis for RQ3; participated in the development of the influence tactics framework; and contributed to writing and revision of the manuscript.\\
\\
All authors read and approved the final version of the manuscript.

\vspace{-1.5em}

\subsection{Data Availability Statement}
\vspace{-1em}
All scripts, datasets, and supplementary materials are publicly available in our replication package~\cite{influencetactics}.
\footnote{\url{https://osf.io/uxhde/overview?view_only=d507800dd6a6434a8c18f8f4607713ea}}

\subsection{Conflict of Interest}
\vspace{-1em}
The authors declare that they have no conflict of interest.

\vspace{-1.5em}

\subsection{Clinical Trial Number}
\vspace{-1em}
Not applicable.

\bibliographystyle{spmpsci}      % mathematics and physical sciences
\bibliography{software}   % name your BibTeX data base

\clearpage
\section*{Authors and Affiliations}

\author{Alex Deaconu \textbf{.} \and
        Anubhav Gupta \textbf{.} \and
        Manaal Basha \textbf{.} \and
        Nicholas Haydu \textbf{.} \and
        Gema Rodr\'iguez-P\'erez}
        
% optional: nice compact formatting
\begingroup
\setlength{\parindent}{0pt}
\newcommand{\authblock}[3]{% #1 name, #2 affiliation lines, #3 email(s)
  \textbf{#1} #2\\
  \texttt{#3}\par\vspace{0.9\baselineskip}
}

\vspace{1.2\baselineskip}

\authblock{Alex Deaconu}%
{\\ University of British Columbia, Kelowna, BC, Canada}%
{alexdea@student.ubc.ca}

\authblock{Anubhav Gupta}%
{\\ University of British Columbia, Kelowna, BC, Canada}%
{anubhav.gupta@ubc.ca}

\authblock{Manaal Basha}%
{\\ University of British Columbia, Kelowna, BC, Canada}%
{manaals@student.ubc.ca}

\authblock{Nicholas Haydu}%
{\\ University of British Columbia, Kelowna, BC, Canada}%
{nich@student.ubc.ca}

\authblock{Gema Rodr\'iguez-P\'erez}%
{\\ University of British Columbia, Kelowna, BC, Canada}%
{gema.rodriguezperez@ubc.ca}
\endgroup

\end{document}